# Coupled CFD-DEM model for dry powder inhalers simulation: validation and sensitivity analysis for the main model parameters

R. Ponzini[1], R. Da Vià[1], S. Bnà[1], C. Cottini[2] and A. Benassi[2,3*]

1 - CINECA Supercomputing Centre, Via Magnanelli 6/3, 40033 Casalecchio di Reno (Italy)

2 - DP Manufacturing & Innovation dept., Chiesi Farmaceutici SpA, Parma (Italy)

3 - International School for Advanced Studies (SISSA), Trieste (Italy)

*Corresponding author address: Chiesi Farmaceutici S.p.A. Largo Belloli 11A– 43122 Parma (Italy)

email address: a.benassi@chiesi.com phone: +39 05211689162

## Abstract

The use of computational techniques in the design of dry powder inhalers (DPI), as well as in unravelling the complex mechanisms of drug aerosolization, has increased significantly in recent years. Computational fluid dynamics (CFD) is used to study the air flow, inside the DPI, during the patient inspiratory act while discrete element methods (DEM) are used to simulate the dispersion and aerosolization of the drug product powder particles. In this work we discuss the possibility to validate a coupled CFD-DEM model for the NextHaler® DPI device against previously published experimental data. The approximations and assumptions made are deeply discussed. The comparison between computational and experimental results is detailed both for fluid and powder flows. Finally, the potential and possible applications of a calibrated DPI model are discussed as well as the missing elements necessary to achieve a fully quantitatively predictive computational model.

## Keywords

## 1 Introduction

In dry powder inhalers (DPI) a drug product, in the form of a fine powder, is aerosolized by an air swirl generated directly by the patient inspiratory act, exiting the inhaler, the airborne particles reach the patient lungs travelling through its mouth and throat [1–5]. To reach the deep airways the active pharmaceutical ingredient (API) particles must have a characteristic size $< 5\ \mu m$, a powder with such small particles is usually very adhesive/cohesive and tend to form large poly-dispersed agglomerates. Moreover, handling small amounts of such fine, poorly flowable powders could be very difficult, thus the API is usually blended with a coarser excipient working as a carrier [6]. Also carrier-free formulations exist where the API is diluted and pelletized with other fine excipients [7]. The efficiency of an inhaler lies in its ability to break down API agglomerates or force the carrier excipient to release the API particles during the inspiratory act. Upon reaching the inhaler outlet, the API particles not properly de-agglomerated or still blended with coarser excipients will not be able to proceed through the lungs. A commonly adopted way of measuring inhaler performances is the fine particle fraction (FPF), i.e. the ratio of API able to reach the deep airways to the total amount of API initially present in the aerosolized dose. The FPF is usually measured in-vitro, using cascade impactors with multiple filtering stages representing the subsequent branches of the human lungs [8].

The performance of a DPI product can be enhanced and optimized by both modifying the formulation of the drug [9–12] ot its manufacturing process [13–16] and acting on the inhaler geometry [17,18]. In either case the non-linear nature of the fluid dynamics as well as the particle-fluid, particle-particle and particle-device interactions make the problem extremely complex. Computational/numerical approaches are thus necessary to gain insight and predictivity. Computational fluid dynamics (CFD), discrete element methods (DEM) and their coupling are indeed valuable tools in this respect. They have been successfully employed to model powder transport by a swirling flow in several contests, e.g. design of industrial cyclone separators [19] and powder comminution in jet mills [20]. Several reviews have been published on their use for inhaler design and aerosolization understanding [21–25]. DPI modelling is also extremely promising in the contest of personalized medicine, allowing to tailor the device properties on specific patient needs [26].

Some authors used DEM to analyse the API detachment from the carrier particles as a function of the carrier-wall collision dynamics, collision geometry, carrier size and coverage degree [27,28]. Other used CFD-DEM to investigate the breakage of fines agglomerates [29,30] or to study the API detachment from the surface of a single carrier particle [31,32]. More recently multi-scale approaches have been developed, for carrier based formulations, in which CFD-DEM is used to model the fluid and carrier dynamics while the release of API fines from the carrier is accounted by empirical relations determined by detailed single carrier particle simulations [33–35]. General purpose guidelines exist to assess the quality and reliability of modelling and simulations for industrial applications [36,37] including medical devices [38], however a common standard for CFD-DEM simulation of DPI has not yet been established.

In this work we used previously published experimental data on NextHaler® to assess the possibility to validate a CFD-DEM model for the dose aerosolization. We start from the validation of the pure CFD model:

- showing the optimal setup able to fit the experimentally measured fluid behavior at the device outlet, and discussing the approximations made;
- using the CFD simulations to characterize the air flow behavior inside the DPI, a region not directly accessible to imaging tools and probes;
- using lagrangian particle tracing, on top of the CFD results, to study the aerosolization of API fines and compare it with experimental data of dose emission as a function of time.

In the second part of the work we discuss the possibility to validate coupled CFD-DEM simulations to study the carrier aerosolization:

- an extensive study varying the DEM and coupling parameters is presented to assess in which way and to which extent they impact the aerosolization process;
- Carrier dynamics is described and discussed based on the set of parameters best fitting the experimental data;
- the effect of the dose protector on inhalation performances will be investigated as in illustration of a possible applications of the validated CFD-DEM model.

We conclude the work by critically analyzing the missing steps and ingredients towards a fully quantitative prediction of a DPI product behavior and its inhalation performance. Many technical aspects have been confined to the appendices to keep the main text lighter and easily readable even by non-experts in numerical simulations.

## 2 Materials and methods

### 2.1 Reference experimental data

The numerical model presented in this paper has been validated against the already published experimental analysis by Pasquali at al. [39,40] performed on the Nexthaler®, a multi-dose, breath-actuated DPI [41]. The inhaler has been loaded with 10 mg doses of a carrier-based placebo DPI blend constituted by a large lactose carrier, with particle diameter $d$ poly-disperse in the range 212-355 µm, and a finer fraction of micronized lactose surrogating an API with particle size between 1-5 µm.

The experimental data set consists of:

- Pressure drop profiles as a function of time for different air flow ramps;
- Particle Image Velocimetry (PIV) analysis of the air plume at the device outlet using 1 µm oil droplets as a passive tracer;
- Fast camera movies of the particle laden lift-up from the dose cup;
- Fast camera movies of the particle plume at the device outlet during dose emission;
- Intensity obscuration profiles as a function of time for a laser beam at the device outlet crossed by the emitted powder cloud, the intensity obscuration should be proportional to the mass flow rate of outcoming particles. In the work by Merusi et al. [40] micron sized beads of a fluorescent markers have been added to the formulation. Assuming they behave as the fine fraction, having same size and being blended with them on the carrier surface, it is in principle possible to monitor only the fine dispersion by filtering the fluorescent intensity signal. A comparison with the total intensity obscuration profile should then allow to distinguish between the fines aerosolization moment and the subsequent carrier emission.

A further effort to distinguish the carrier emission dynamics, explicitly simulated in this work, from the fine fraction one, has been done by reprocessing the fast camera movies.

- Using standard image processing algorithms, detailed in Appendix A, we have been able to distinguish carrier particles from the fine fraction clouds, to count them and to estimate their size and shape. An emission curve for the sole carrier component can thus be plotted as a function of time.
- Finally, we have been able to track certain carrier particles through the movie frames and to estimate their velocity by the frame rate. Details are also illustrated in Appendix A.

During the experimental measurements the flow rate profiles, applied to the device through a pump, have an initial ramp of variable duration $t_{ramp}$ up to the maximum value $Q_{max}$, which is then kept constant for the rest of the recording of total time $t_{tot} = 1.4$ s. Different solutions have been tested combining $t_{ramp} = 0.3, 0.7$ and $1.2$ s with $Q_{max} = 40, 60$ and $80$ l/min. Unfortunately, the exact behaviour of the flow rate during the initial ramp has never been directly measured, thus limiting the possibility to impose it as boundary condition into the numerical model as it will be fully discussed later.

## 2.2 Numerical methods

The simulations have been performed on a modular test rig geometry rather than on the Nexthaler® one. However such test rig has been designed to have the same identical internal volumes and geometry of Nexthaler® and the same aerodynamic resistance [17,42], thus for our purpose the two are perfectly equivalent. The Nexthaler® incorporates a dose protector which initially covers the dosing cup containing the 10 mg of formulation to be aerosolized. It is designed to retract only when the pressure drop, between inlet and outlet, generated by the inspiratory act of the patient reaches a pre-set trigger value of 2 kPa. In this way the powder dose is exposed to an already developed air swirl, thereby ensuring its complete disaggregation, fluidization and an efficient dose emission. The test rig does not contain a dose protector, and in any case at present we are not interested in a detailed modelling of the dose protector motion during the simulations. It is however reasonable to assume the dose protector motion does not influence the formation of the air vortex and the development of the fluid structures during the aerosolization. The same cannot be said from the powder aerosolization perspective: the exposition of the powder to the low velocity fluid, while the air swirl is not yet formed, can lead to very different powder dispersion results compared to the case of a sudden exposition to an already formed and high energy swirl. On the other hand, fast camera movies of the powder cup during the device actuation revealed that the cup is not perfectly sealed by the dose protector and the powder dose is waved, by the fluid, even before the latter is triggered [39]. Finally, it must be noted that the dose protector opening is not instantaneous, taking $3-4$ ms to be completed, and the current resolution of the fast camera movies is not able to accurately resolve its motion. To conclude, the explicit and exact simulation of the dose protector opening would require a detailed analysis which is beyond the scope of this first general work, requiring also further ad-hoc experimental observations; therefore we decided to simulate the two following cases:

- an aerosolization without the dose protector, with the powder exposed to the air swirl since the very beginning of the fluid motion;
- expose the powder to the fluid only when the dose protector is supposed to open, i.e. when a 2 kPa pressure drop is reached in the fluid simulation.

These two extreme cases should bracket the realistic powder behaviour. Moreover, the first case is easily implementable in real experimental conditions, by pre-triggering the dose protector manually, and it is the standard way the test rig is supposed to work. It is thus of our interest to compare the differences in the two aerosolization mechanisms, i.e. with and without the intervention of the dose protector.

A sketch of the employed test rig geometry is shown in Figure 1 (a), while panel (b) of the same figure shows the simulated air volume comprising the rig internal volume and a *plenum* just outside the rig outlet. The plenum has the same size of the discharge chamber in the experimental setup [39] and t is necessary for the air/particle plume to develop and being imaged by the PIV equipment.

The air and drug product flows, as well as their interaction, have been simulated through a 4-ways coupled CFD-DEM model. Numerical calculations have been performed through the CFDEM® Coupling software [43,44] from DCS Computing GmbH [45]. The software couples modified CFD solvers from OpenFoam® [46,47], for fluid flow modelling, to the DEM software Aspherix® [48], for powder flow modelling.

### 2.2.1 CFD simulations

For the fluid dynamics a modified version of the time-dependent, incompressible, pressure-based solver *pisoFoam* has been adopted. Compressible solvers fully coupled with the solid phase, and able to correctly describe the multi-phase transient, are currently unavailable in CFDEM® Coupling. From a pure CFD perspective, the Mach number $Ma$ can be estimated to assess how acceptable is the incompressible flow assumption, as a rule of thumb $Ma = 0.3$ is considered the limit above which the gas compressibility cannot be neglected. As illustrated in the next sections, the maximum fluid velocity recorded during the simulations is $v^f_{max} = 95$ m/s, a worst-case value found only in the $Q_{max} = 80$ l/min case. Using the speed of sound of air at ambient conditions $c = 343$ m/s one gets $Ma = v^f_{max}/c = 0.27$, even in the worst case we are always below the $Ma = 0.3$ limit. From the point of view of the fluid-particle coupling things can be more complex and estimating *a priori* the impact of the incompressible flow assumption is impossible. After setting and validating our coupled, incompressible model, we performed a pure CFD steady state calculation, relaxing the incompressibility assumption, using the OpenFoam® solver *rhoSimpleFoam*. This allowed us to evaluate *a posteriori* the implications of neglecting air compressibility, results are presented in Appendix B.

In a cylindrical pipe a fluid is known to become turbulent at Reynolds number $Re^f$ equal to 2500 [49]. Using as characteristic system lengths the inlets width $\ell = 1.5$ mm or the pipe diameter $D = 7$ mm one can estimate the maximum Reynolds number $Re^f = 10000 \div 43000$. With such values the flow is clearly turbulent (the same $v^f_{max}$ of the $Ma$ estimation has been adopted). Turbulence is included in the calculations through the effective eddy-viscosity model $k\omega - SST$ [50]. As deeply discussed in the inhaler modelling literature such choice represents a good compromise between accurate flow description, including its transitional regime, and computational cost. The fluid behaviour close to the boundaries is modelled through the low Reynolds number wall functions [51].

Hexahedral-dominant meshes have been constructed for the discretization of the integration volume using the standard meshing tools of OpenFoam®, namely *blockMesh* and *snappyHexMesh*. Close to the walls the meshes have been finished with surface layers, finer elements following closely the geometry profiles, to improve the description of the fluid velocity decay inside the physical boundary layer. The mesh boundary layer thickness has been chosen according to the flat plate physical boundary layer formula [49]: estimating a meaningful average reference velocity and targeting a first cell height of about 5, in terms of the dimensionless length usually referred to as $y^+$ [52]. The reliability of this estimation has been verified *a posteriori* by monitoring the $y^+$ during the simulations and ensuring that its maximum average value did not exceeded a value of 10, according to the low Reynolds wall function model prescriptions. A mesh sensitivity study has been conducted to determine the minimum size of the mesh producing well converged numerical results independent from the mesh cell size. Meshes ranging from 500K to 2.6M elements have been generated and evaluated. In the mesh generation process only the reference mesh cell size has been changed, keeping constant the relative refinement ratios and the boundary layer cell size. The selected mesh, containing roughly 900 thousand elements, is shown in Figure 2 (a): the mesh is finer approaching the walls and coarser in the plenum above the device outlet, the insets show how 5 or 3 surface layers have been applied to the pipe/outlet and to the swirl chamber walls respectively. The average characteristic length of the mesh elements $\overline{\Delta x}$ is $372$ µm, while the smallest element has a characteristic size $\Delta x_{min} = 36$ µm. Since the smallest elements have low aspect ratios their thickness can reach roughly $25$ µm. Pressure drops between the inlet and the outlet, as well as some average velocity value, have been monitored during the mesh sensitivity analysis (a flow rate driven condition has been adopted in this case); their values are plotted in Figure 2 (b) showing weak or absent dependence on the mesh size.

The air flow through the device is controlled by the application of time dependent boundary conditions. While in the experiments a controlled flow ramp is applied through a pump downstream the device, measuring the resulting inlet/outlet pressure drop in time, in the simulation we preferred to control the air flow by imposing the inlet/outlet pressure drop while monitoring the freely developing flow rate. The motivations for such choice are deeply illustrated in Appendix B, together with a comparison between the results obtained with the two alternative driving mechanisms. In the adopted *pressure driven scheme* the applied boundary conditions are summarized in Table 1 ($k$, $\omega$ and $\nu_t$ are scalar fields necessary to treat turbulence: the turbulence kinetic energy per unit mass, the specific rate of turbulent energy dissipation and the kinematic eddy viscosity respectively). As initial condition the fluid has been set at rest at ambient pressure in every point of the device inner volume. The initial conditions for the other fields are summarized in Table 2.

From the point of view of the pure CFD the reference number to look at, while setting the simulation time-step, is the fluid Courant number $Co^f$ or the ratio between the time integration step $\Delta t_{CFD}$ and the typical time taken by the fluid to cross a discretized volume element. To ensure that the temporal discretization of the fluid dynamics is appropriate for the chosen spatial discretization, this number must be smaller than or closer to 1. Using for the fluid velocity the worst case represented by $v^f_{max}$ and the average mesh element size $\overline{\Delta x}$ the Courant number becomes $Co^f = v^f_{max}\,\Delta t_{CFD}/\overline{\Delta x}$, a good choice for the time-step in order to set $Co^f \cong 0.5$ is $\Delta t_{CFD} = 2$ µs.

### 2.2.2 DEM simulations for carrier particles

As a first rough approximation the dose powder has been represented by spherical, poly-disperse particles whose size distribution corresponds to the real carrier one, while the lactose fine particles mimicking the API inhalable fraction are not explicitly simulated. This is due to some intrinsic limitations of the DEM technique itself which make heavily poly-disperse systems very expensive to treat from a computational point of view:

- The time integration step $\Delta t_{DEM}$ is limited by the diameter $d$ of the particles to be simulated. As a rule of thumb $\Delta t_{DEM}$~20% of both Rayleigh and Hertz time to ensure a good resolution of each collision event [53]. To fulfil this condition when simulating $d = 1\,\mu m$ particles one should set $\Delta t_{DEM} < 10^{-9} s$ which makes very hard to achieve a simulated time of 1 second in a reasonable amount of real time.
- A possible way to mitigate this time resolution problem is to choose the Young modulus for both particles and walls to be orders of magnitude smaller than the real lactose and plastic ones, this keeps the Rayleigh time small thus allowing for large integration time-steps. Being less stiff both particles and walls will experience larger deformations, but the same particle-particle and particle-wall forces will be generated during the collisions. This is a standard trick applied in most of the DEM simulations and it is known not to affect the results as long as the simulated powder does not undergo a strong compression, like e.g. in tableting simulations or ball milling simulations [54].

- Coarse carrier and fine API particles, whose diameters differ by 2 orders of magnitude, end up travelling with very different velocities while dragged by the fluid. Thus, without a $\Delta t_{DEM}$ fine enough, large and small particles could inter-penetrate each other resulting in numerical instabilities or in missing some of the particle-particle collisions at all.
- Other issues emerging from a strong poly-dispersion are of pure numerical nature. For instance, the standard, multi-purpose, algorithms for the update of the neighbor lists of each particle can become extremely inefficient. The same happens to the algorithms distributing the computational load among different cores in parallel computations. This could result in a dramatic increase of the computational time compared to the same simulation run with the same mass of mono-disperse powder. Some of these issues have been addressed in the literature [55] but are not yet implemented in most of the available DEM codes.
- Even with all the ad-hoc optimized algorithms in place, an explicit simulation of the fine API particles remains prohibitive: assuming only few percent w/w of API concentration, with fine API particles of 1-5 μm diameter, would result in many millions to tens of millions of particles. These numbers are still affordable through high performance computing resources, for proof of principle calculations, but out of reach for engineering service applications, such as modelling campaigns for the design/optimization of specific DPI devices. Limiting our simulations to 10mg of carrier particles only means working with $\sim 800$ poly-disperse particles or $\sim 1250$ mono-disperse particles with $d = 100$ μm. These numbers are easily handled without the needs of significant supercomputing resources.
- If, from one hand, the use of an eddy-viscosity approximation allows to model the correct energy dissipated by the turbulent air flow at a reasonable computational effort, on the other hand it leaves us with a fluid velocity field smoothened from the turbulent eddies fluctuations, not explicitly simulated. This is not supposed to be a problem for the large carrier particles: given their large mass, their motion will be dominated by their own inertia being not, or only very weakly, disturbed by the turbulent eddies. This attitude is quantified by the Stokes number $Stk$, i.e. the ratio between the characteristic time necessary to accelerate particles and the characteristic time of the fluid motion. The time necessary to reach the terminal velocity is typically used as the characteristic relaxation time for the particle; if, for the evaluation of the fluid time, the ratio between the fluid average velocity and the characteristic system size is used we are left with $Stk = \rho_p d^2 v_{max}^f / 18\, \mu\, D$. Using the diameters of the carrier particles leads to values in the range $1200 \div 13000 \gg 1$, whereas for 1 µm particle diameter one gets $0.06 \ll 1$, i.e. the micron sized particles will follow strictly the air streamlines. As the latter do not incorporate the effect of the turbulent eddies, the micron sized particle scattering and distribution will be completely meaningless. A possible way to account for the small size particle dispersion by the turbulent eddies is to incorporate in the DEM simulation a stochastic force whose intensity and spatial correlation depends on the turbulent energy density obtained by the CFD calculations [22,56]. Of course, with such “statistical” approach only the average and the integral properties derived from the particle trajectories are meaningful. Such stochastic force models are not currently implemented in the software we used, which makes us once more unable to explicitly simulate the API fine particles deposited on the carrier surface.

The presence of the fine lactose/API particles, covering the carrier surface, can still be implicitly included through the inter-particle cohesion energy, at list to a qualitative extent. In fact, small amounts of fines result in free-flowing DPI mixtures, easily to aerosolize while, increasing the fines content, the mixture becomes highly cohesive, poorly flowing and thus hard to disaggregate and aerosolize. However, the detachment of the API particles from the carrier surface, their disaggregation and dispersion crossing the inhaler up to the outlet cannot be directly studied with the current model. As we will see many considerations can still be drawn based on the behavior of the carrier particles and the turbulence properties of the air flow.

The DEM particle density has been set to the lactose one $\rho_p = 1525\ Kg/m^3$. For both the particle-particle and particle-wall interactions the Hertz-Mindlin model has been selected, a contact history for the tangential force component has also been included [57]. A constant direction torque model has been used for the rolling friction [58], thus both a sliding $\mu_s$ and a rolling $\mu_r$ friction coefficient must be specified for particle-particle and particle-walls contacts. We have many evidences that our carrier powder has a free-flowing character, each particle moving independently from the others, this means volume forces, i.e. gravity, prevail significantly over surface forces, i.e. adhesion/cohesion. On the other hand, even a simple inter-particle cohesion model, such as the Johnson-Kendall-Roberts model [59], requires the introduction of a cohesion energy density parameter that we cannot measure experimentally. Thus, in this preliminary work, we decided to neglect any particle-particle and particle-wall cohesion force, keeping the DEM simulation as simple as possible and avoiding the proliferation of tunable model parameters. Adhesion/cohesion effects can be easily reintroduced, to treat high payload carrier-based formulations or alternative carrier-free formulations, if specific data are available to calibrate their model parameters. The choice of the other DEM model parameters, as well as their impact on the aerosolization process, will be discussed in detail in the results section.

The gravitational force in both CFD and DEM simulation is switched on and points along the negative direction of the z-axis (with respect to Figure 1). This is because in the reference experiments as well as in the in-vitro testing of inhalation performances, the device is supposed to be tilted 90 degrees with the pipe horizontal and perpendicular to the gravity direction.

### 2.2.3 CFD-DEM coupling

The large number of particles and their small size require the use of an *unresolved* coupling for the description of the fluid-powder interaction, with the CFD mesh elements larger than the particle diameter and able to contain, in principle, many particles. The particle properties necessary to compute the interaction forces and the coupling term result from the average over all the particles laying in the same CFD mesh element. The aerosolization of a small dose of dry powder is an unsteady, fast phenomenon involving turbulent swirling air flows. In such conditions, according to the literature prescriptions, the so-called *model A* coupling is necessary as the fluid phase moving through the powder mass cannot be considered steady and uniform [53]. Details of the coupling model implementation can be found in the reference paper by Zhou et al. [60].

The most relevant term in the powder-fluid interaction is certainly the drag force. Many different drag models exist with empirical correlations fir the drag coefficient fitted from experimental data or derived from numerical simulations [53]. In our specific application the powder is initially concentrated in the dosing cup with very small fluid volume fraction values of $n = 0.44$, during the early moments of the lift-up the powder remains very dense. As the swirl strengthen and the aerosolization proceeds, the powder flow shifts from a contact-dominated, to a collision-dominated, to a collision-free regime, with $n$ gradually reaching 1. Thus, the drag model of choice must be valid in both the dense ($n < 0.8$) and dilute ($n \geq 0.8$) condition. Moreover, most of the drag models employed in the literature have been fitted and validated in a specific range of particle Reynolds numbers $Re^p = n\, \rho_f d\, v_{slip}/\mu$, where $v_{slip} = \left|\overrightarrow{v^f} - \overrightarrow{v^p}\right|$ is the modulus of the slip velocity, i.e. the difference between particle and fluid velocity. From our simulations the maximum slip velocity reaches the value $v_{max}^{slip} = 40$ m/s with a maximum $Re^p$ of 1200. The two criteria just illustrated limit our selection of drag models to the following three among all the implemented ones: Di Felice [61], Gidaspow [61] and Koch-Hill [62]. While the first two models should work fine for any range of $n$ and $Re^p$, the Koch-Hill one has been tested only for $Re^p \leq 50$ thus could not be perfectly suited to describe the drag force during the late phase of aerosolization, when the particles accelerate strongly along the pipe. All these models have also some other major limitations:

- They have been designed for mono-disperse particles. Extensions exist in this respect [53] but, to the best of our knowledge, they have never been tested on strongly poly-disperse systems having two orders of magnitude difference in particle diameter such as carrier based DPI. This is another significant limitation to be overcome before a direct DEM simulation of both carrier and API fines can be considered reliable.
- They have been designed for spherical particles. Also in this respect extensions exist to include both the surface roughness of particles and their deviation from sphericity [53,63]. With the long-term aim to reach a quantitative agreement between simulations and experiments, the suitability of these models should be tested on the typical carrier particles used in DPI formulations, i.e. crystals of lactose, mannitol or other sugars whose shape and surface are certainly very irregular.

- They are not adequate to treat particles attached to or in the close vicinity of a wall, where the boundary layer causes the fluid velocity to quickly drop to zero. This is particularly important in the shear aerosolization of powder particles, as initially they usually lay at rest on a surface before being lifted and scattered in the fluid volume. To this aim the DEM code itself should be able to discriminate between those particle far enough from walls to be out the boundary layer influence and those not, and apply to the latter a different empirical correlation such as the one developed by Sweeney and Finlay [64], for particle in close contact, or the Brenner one [22], for particle in the vicinity of a wall. In any case these correlations should be generalized to the presence of a dense system of particles deposited on surfaces (small $n$ values).

Lift forces could also be important to correctly describe the carrier dispersion, especially in the initial moments. For the shear lift, i.e. the lift force provoked by a fluid velocity gradient, we tested the empirical correlation by Mei [65,66]. The Magnus lift force, generated by particle spinning, could also play a role however it is not currently implemented in the simulation software. The same considerations on drag forces above apply when particles are attached or close to the device walls: proper correlations for the lift coefficient are available in literature for the case of single particles but not for dense systems [64,67,68].

Besides drag and lift forces our coupling model also accounts for the particle-based pressure gradients forces as well as the particle-based viscous forces, i.e. the forces generated by spatial inhomogeneities in the viscous stress tensor [60]. Finally, the Archimedes force, the added mass force and the Basset force have been intentionally neglected given the small density of the fluid or the large discrepancy of many orders of magnitude between the fluid and solid density.

It must be mentioned that the constraint of having $\Delta x > 2 \div 3\ d$, typical of plain unresolved CFD-DEM coupling [69], is not met in the proximity of the walls. This is due to the mesh refinement that is necessary there to correctly resolve the boundary layer. The same is true, everywhere in the CFD integration domain, for the largest particles included in our poly-disperse DEM distribution. This situation might cause numerical instability due to abrupt discontinuities in the coupling fields, one known way to deal with the problem is to introduce a smoothening of the coupling field around each particle. To achieve this, while conserving the smoothened quantity, it is possible to use a simple diffusion equation [70,71]. The diffusion coefficient is calculated as ${l_s}^2/\Delta t_{CFD}$ where $l_s$ is a characteristic smoothening length having of course the property $l_s \geq d$. In our specific case we have set $l_s = 240\ \mu m$.

In a plain DEM simulation of monodisperse spherical particles the time-step choice is usually driven by the Hertz and Rayleigh times [53]. Taking $\Delta t_{DEM}$ to be roughly $20\%$ of both these reference times should ensure the particle deformations and collisions to be accurately resolved. The Rayleigh time depends solely on the size and elastic properties of particles, with the chosen values (illustrated in Table 3) its value lies in the interval $0.4 \div 1.3 \times 10^{-5}$ s; the Hertz time depends also on the maximum particle velocity $v_{max}^p$, in our typical simulations (e.g. $Q_{max} = 60$ l/min with $t_{ramp} = 0.3$ s) $v_{max}^p = 10$ m/s thus setting the Hertz time in the range $0.8 \div 2.5 \times 10^{-5}$ s. A value of $\Delta t_{DEM} = 0.5$ µs would be already enough to fulfil the previous criteria, to stay on the safe side we have chosen $\Delta t_{DEM} = 0.2$ µs so that a factor $10$ exists between the DEM and CFD time-steps. Having now a coupled system other factors might intervene in the definition of the fluid and particle time integration steps and their relative proportion. Analogously to what has been done for the fluid, it is possible to define the ratio between the CFD time integration step and the time taken by a particle to cross a discretized volume element, such ratio is called particle Courant number $Co^p = v_{max}^p\, \Delta t_{CFD}/\overline{\Delta x}$. For a particle to perceive fluid velocity variations and to capture correctly its exchange of momentum with the fluid, one should run the simulations in the condition $Co^p \leq 1$. With our typical numbers, and with our choice of $\Delta t_{CFD}$, we got $Co^p \cong 0.05$ or a higher value of 0.7 if the average discretization size $\overline{\Delta x}$ is replaced by its minimum $\Delta x_{min}$. A last quantity to look at, in making sure we made a correct choice for $\Delta t_{CFD}$ and $\Delta t_{DEM}$, are the characteristic relaxation times for particle $\tau_p$ and fluid $\tau_f$, i.e. the time necessary for them to exchange energy/momentum and reach a steady condition. From the perspective of a single particle $\tau_p = \rho_p d^2/18\,\mu$ [22], with our range of particle diameters this means $0.09 \div 0.26$ s; from the fluid point of view the effort in dragging and accelerating particles depends on the particle density and thus on the fluid volume fraction $\tau_f = \tau_p \ \rho_f/\rho_p \, n/(1-n)$, in the initial moments of aerosolization, when $n$ is large we get $\tau_f = 5.6 \times 10^{-5}$ s. In both cases the relaxation time is much larger than the respective integration time steps, ensuring a good time resolution for both the dynamics and their coupling.

The relevant material properties characterizing the air fluid and the particle mechanical response are listed in Table 3.

### 2.2.4 Lagrangian tracing for API particles

As mentioned in the previous sections the very small Stokes number of the API fines makes them complex to be treated within a CFD-DEM coupling scheme. However, to have a first qualitative description of their aerosolization, lagrangian tracing featuring hard-sphere collisions could be suitable. In this work we used the function object icoUncoupledKinematicCloud available in OpenFoam® to follow the trajectory of fine spherical lactose particles, released in the cup volume during the initial transient, up to the device outlet and estimate their residence time. In such simplified picture only gravity and drag forces have been considered using the empirical correlation by Schiller and Naumann [53]. In the hard-sphere model the collisions are instantaneous, this allows to simulate micron-size particles without the need of nano-size time steps, for our calculations we used $\Delta t_{LT} = 5 \times 10^{-6}$ s. The function object allows also to include the effect of the turbulent eddies on the particle trajectories via a stochastic eddy interaction model [72]. Given the very small number of particles inserted both the particle-particle collisions and the effect of the particles on the fluid have been neglected, having thus a passive (1-way) lagrangian tracing. With such a simple model the only parameters to be calibrated are the static friction coefficient $\mu_s$ and the restitution coefficient $e$ for particle-wall collisions.

## 3 Results and discussion

In the first part of this section we illustrate in detail the comparison between CFD simulations and experimental measurements. We start showing how, thanks to the proper selection of mesh, initial and boundary conditions, it is possible to achieve a quantitative matching for the simulated flow rate and velocity profiles, as well as a good qualitative comparison with the 2D PIV maps. Once the CFD model has been validated we use it to describe the fluid behavior inside the device and to study the behavior of the API fine particles in the first stages of their aerosolization. Subsequently we proceed with the calibration of the carrier DEM model, and its coupling parameters, against the experimental time dependent intensity profiles describing the carrier emission from the device. Finally, with the coupled CFD-DEM model calibrated, we come back to the aerosolization physics, considering explicit the carrier particle dynamics.

## 3.1 Validation of the CFD model

As anticipated in section 2.2 the air flow is generated applying a time dependent inlet/outlet pressure drop as measured in the reference experiment. As a first validation step we verified that the velocity profiles at the outlet (averaged along line 1 and 2 of Figure 2) corresponds with those measured by PIV experiments as a function of time. An example of such comparison is given in Figure 3 (a), the average values are indeed in agreement. Notice how the experimental curve is much noisier than the simulate ones, this is due to the turbulent eddies not explicitly simulated in our Reynolds-averaged eddy-viscosity approach. We also performed a CFD simulation without including the turbulent dissipation, i.e. forcing a laminar behavior for the fluid, this led to an overestimation of the fluid velocity and to more pronounced time fluctuations. This was expected due to the artificial and non-physical reduction of the energy dissipation. Afterwards, we verified that the proper flow rate $Q_{max}$ is reached upon application of the corresponding inlet/outlet pressure drop $\Delta p$, four cases have been plotted in Figure 3 (b) corresponding to those illustrated in the experimental reference works. The correct flow rate onset confirms that the simulated geometry (the test rig one) has the same aerodynamics resistance of the real Nexthaler®.

A comparison between CFD and experimental data can be also made with the 2D velocity maps from the PIV experiments. Results are shown for the separate in-plane components of the air velocity in Figure 4 and for the air velocity magnitude at different time instants in Figure 5. From a general point of view the experimental velocity maps look blurrier than the simulated ones, this is due to the presence of the turbulent eddies decomposing the main flux structures. In our Reynolds-averaged approach to turbulence the multi-scale eddies are absent thus the simulated flux structures remain more coherent in time and space and have more clear boundaries and gradients. A time average can also be performed on the experimental data, smoothening out the “noise” generated in the PIV images by turbulence. This average is shown in the last map of row (a) and (c) of Figure 5, indeed it looks much similar to the simulated maps, both the instantaneous and the averaged ones. The y-component map of the velocity field represents a cross section of the vortex structure emerging from the device outlet, it reveals how the air direction points outwards the device (yellow and orange plumes) only near the vortex periphery, while in the vortex core the air flows inside the device (blue region). The agreement between PIV data and simulations is quite good with the orange plumes having the same thickness, aperture angle and average length, although some instantaneous PIV snapshots might show shorter plumes due to their decay into turbulent eddies. The vortex structure is not steady in time, it oscillates with a characteristic frequency transferring part of its energy from y to x direction. The x-component map shows in fact a shedding of high-speed spots of alternate sign moving along the vortex periphery. The velocity magnitude images of Figure 5 reveal a good agreement between simulation and experiments also in the time evolution of the swirling flow. Air flows initially out of the device as a uniform column that splits in two halves forming perfectly vertical plumes, the plumes subsequently bend as the vortex widens and increases its intensity up to its steady state shape.

### 3.2 Internal fluid behavior based on pure CFD model

Pressure drops and velocity maps recorded at the device outlet indicate a good qualitative and quantitative agreement between simulation and measured data both in the spatial arrangement of the flux and in its time behavior, validating our model at least from the pure CFD side. We can now use it to visualize what happens inside the device, along the pipe and in the swirl chamber, near the cup hosting the drug powder dose to be aerosolized, i.e. in those regions where no direct experimental measures are currently available or where they cannot be recorded owing to technical difficulties.

The behavior of the fluid in the initial moments of the transient is captured by the time sequence of Figure 6 (a) for the case $Q_{max} = 60$ l/min and $t_{ramp} = 0.3$ s. The streamlines entanglement reveals the formation of two similar vortexes rotating and enveloping on each other to form a unique stable swirl structure around $13 - 15$ ms. From that moment onwards, both the pressure drop and the air velocity increase significantly but the shape of the fluid structure does not change anymore, notice how the stabilization of the vortex occurs much earlier than the dose protector release. The last two images of panel (a) show the final structure of the vortex at the end of the initial transient. Panel (b) and (c) of Figure 6 give a picture of the pressure distribution and the velocity field once the steady state is reached, from $300$ ms onwards. The CFD simulations reveal how air experiences two strong accelerations while entering the swirl chamber, through the two inlets, and while exiting the swirl chamber due to the sharp edge in the connection with the pipe (close to line C). A weak precession motion of the air vortex is also visible which generates local time fluctuations in the velocity field, these become significant at the sharp edge between swirl chamber and pipe and propagate all along the pipe through the device outlet. The largest pressure difference is recorded at the bottom of the swirl chamber, where the cup is located. In the terminal part of the pipe and at the outlet a smaller pressure gradient onsets and air decelerates. Notice how, in agreement with the PIV data, air flows inside the device in the vortex core at the device outlet (e.g. across the A line). Proceeding upstream along the pipe such blowback effect disappears, already close to line B the velocity in the core vortex is small but positive. The tangential component of air velocity has the typical anti-symmetric profile all along the y-axis, confirming the swirling nature of the flow. The radial component is almost everywhere pointing inwards the vortex, dragging the particles towards its core and opposing the centrifugal force, pushing them against the inhaler walls. The phenomenology just illustrated remains unchanged even when different $Q_{max}$ and $t_{ramp}$ are applied, they only affect the maximum intensity of the swirling flow and the time necessary to achieve it.

A last point to be addressed, in assessing the quality of our simulations, is the correct resolution of all the time dependent fluctuations occurring in the fluid dynamics due to the air vortex precession motion. The transient time $t_{ramp} = 0.3 \div 1.2$ s is always much larger than our time resolution $\Delta t_{CFD} = 2$ μs, however, due to the flow separation at the sharp edges of the swirl chamber, shedding occurs and the air vortex fluctuates in time both inside and outside the pipe. These fluctuations are clearly visible in the PIV experimental data and are also present in our simulations. Figure 7 (a) shows, as an example, the velocity profiles as a function of time recorded in three different points of the inhaler pipe and outlet. Regular time fluctuations are clearly visible whose frequency is different depending on the recording position. Fourier transforming such time signals allows us to have a frequency spectrum of the velocity field oscillations, this is presented in the same panel. The shortest characteristic fluctuation time is about 0.6 ms, still much larger than $\Delta t_{CFD}$. It is interesting to notice how, switching off the turbulent dissipation, i.e. forcing the air flow to be laminar, the velocity profiles look noisier and more irregular, different frequencies appear in the spectrum, see panel (b) of Figure 7. However, even in this unphysical condition, the shortest characteristic oscillation time remains above 0.25 ms $\gg \Delta t_{CFD}$. Panel (c) shows the velocity profiles extracted from the PIV data, unfortunately the sampling rate of 1kHz is not enough to achieve the interesting frequency range between 5 and 20 kHz in the Fourier transform, a direct comparison with the real time fluctuation scales is thus not possible. It must be said that, even with higher frequency PIV data, it might be difficult to extract the shedding frequencies of the vortex and discriminate them from the time scales of the turbulent eddies.

We conclude the section by analysing the intrinsic relaxation time of the system $\tau^*$: so far we have switched on the air flow through different ramps of duration $t_{ramp}$, which is, instead, the characteristic time for the system to react to an instantaneous perturbation? How does it compare with $t_{ramp}$? To answer this questions we performed different CFD simulations switching on instantaneously constant pressure drops $\Delta p$ corresponding to different $Q_{max}$ and recorded the flow rate profiles. Few examples are given in panel (a) of Figure 8 (continuous lines) such oscillating profiles can be fitted with the following equation:

$$Q(t) = Q_{max}\left[\tanh\left(\frac{t}{\tau^*}\right) + e^{-t/\tau^*}\sum_{i=1}^{3} a_i \sin\left(\omega_i\right)\right] \qquad (1)$$

$\tau^*$ being the characteristic relaxation time of the system, $a_i$ and $\omega_i$ are other fitting parameters. The fits are shown in the same panel with dashed lines. Like the aerodynamic resistance, $\tau^*$ depends on the specific device geometry, it also depends on the maximum induced flow rate $Q_{max}$. Such dependence is highlighted in Figure 8 (b) and seems to be almost perfectly linear. Notice how $\tau^* \ll t_{ramp}$ independently from $Q_{max}$: this explains why the flow rate profiles of Figure 3 and Figure B.2 grow smoothly without the sinusoidal fluctuations exhibited in Figure 8 (a).

### 3.3 Considerations on aerosolization based on the pure CFD model

Assuming the API particles to be lodged on the carrier surface, filling the cup in the bottom of the swirl chamber, some considerations can be made on the nature of the air flow lifting them up and entraining them into the forming swirl during the initial transient. Although we already discussed in the previous sections the complex 3D nature of the swirling flow, for the sake of simplicity and for a rough estimation of the orders of magnitude of the forces at play, we will assume a 1D straight air flow from the inlet to the cup, see Figure 9 (a). The API particles invested by such flow are at rest on the carrier surface on which a boundary layer will develop, as depicted in Figure 9 (c), decreasing the air velocity from the free stream value $v_\infty$ to 0. The flat plate theory can be used to estimate the velocity profile $v(x,y)$ on the bottom of the swirl chamber and on the carrier surface [1]:

$$v(x,y,t) = v_\infty(t)\left[\frac{3}{2}\frac{y}{0.918\,\delta(t)} - \frac{1}{2}\left(\frac{y}{0.918\,\delta(t)}\right)^3\right] \qquad (2)$$

being $\delta = 5\,x\,\sqrt{Re_x}$ the boundary layer thickness and $Re_x = v_\infty\,x\,\rho_f\,/\mu$ the local Reynolds number, $v_\infty$(t) is instead estimated from $Q_{max}$ knowing the inlets area. Figure 10 (a) shows such profiles along $y$ for three different $x$ values for the case $Q_{max} = 60$ l/min: at the cup centre (blue), halfway from between the inlet and the cup (yellow) and 1/4 of the inlet-cup distance (green). The yellow dashed line represents the value of the velocity magnitude extracted from the 3D flow CFD simulations at 1/2 of the inlet-cup distance, despite the strong assumption of one-dimensional flow the estimated profile compares nicely with the simulated one, the order of magnitude is indeed correct. Once the steady flow is reached $\delta$ is around 240 μm, i.e. two orders of magnitude larger than the API particle diameter, it is thus reasonable to assume that such fine particles experience drag and lift forces from a laminar flow. From the 3D CFD simulations a plot of $y^+$ can be extracted at a given distance from the swirl chamber bottom, when its value is < 5 the air flow is laminar. The inset in panel (b) of Figure 10 shows a $y^+$ map calculated at 1.4 μm from the bottom, i.e. useful for particles of 3 μm of diameter, indeed the flow is always laminar everywhere, confirming again the estimation made with the 1D flat plate model.

The flat plate theory is formally correct only when applied to a steady state condition however, as shown in the previous section, the response time of the air flow to instantaneous perturbations is of few ms, while our applied flow ramps are of 300 ms. We can thus reasonably assume our inhalation to be quasi-static and the flow to readjust almost instantaneously while we rise $Q(t)$. Replacing $Q_{max}$ with Q(t) equation (2) becomes time dependent and can be used to estimate the air velocity at 1.5 $\mu m$ height, and thus the drag and lift experienced by a $d = 3\ \mu m$ API particle. The local Reynolds number as well as the boundary layer thickness as a function of time, in the centre of the cup, are shown in panels (b) and (c) of Figure 10 for $t_{ramp} = 0.3$ s and $Q_{max} = 60$ l/min. Panel (d) of the same figure shows the drag and lift forces calculated in the laminar approximation. A comparison with the typical values of gravitational and adhesion forces for lactose particles, reported in the literature [1], reveals how both lift and drag forces could detach API fines from the carrier well before the dose protector shift (vertical dashed line at 45 ms). This simple calculation allows us to release the fine particles at 45 ms during the simulation and follow their aerosolization using the passive lagrangian tracer described in section 2.

We start by releasing 1000 particles with $d = 3\ \mu m$ representative of a typical API fine fraction with $1 \div 5\ \mu m$ diameter. As insertion region we tried both the full cup volume and the 2D circular region separating the cup volume from the swirl chamber only, no significant difference has been noted in the particle behavior and in the calculated residence time. The particle trajectories are almost completely unaffected by the choice of $\mu_s$, we have thus set the intermediate value $\mu_s = 0.5$. On the contrary the restitution coefficient $e$, i.e. the ratio between the particle velocity after and before a collision, plays a major role: we tested several collision conditions ranging from perfectly elastic ($e = 1$) to almost completely inelastic ($e = 0.05$). The results are shown in Figure 11 (a) displaying the normalized mass flow rate at the device outlet, the residence time being estimated through the position of the main peak. The larger $e$ the faster is the particle emission, only with $e < 0.15$ the emission peak position start to be independent of $e$ and to approach the measured values (black dots). Elastic particles preserve more kinetic energy during the first collisions with the bottom and lower side walls of the swirl chamber, in the early stage of aerosolization, this allows them to populate the central part of the chamber where the swirling flow is more energetic and lift them more efficiently.

Inelastic particles remain more adherent to the walls where the swirling flow is less developed preventing them to accelerate significantly. This behaviour is illustrated in the snapshots of Figure 11 (b) comparing the particle trajectories for the two extreme cases $e = 0.05$ and $1.0$ in two subsequent instants of the early aerosolization stage. Upon reaching the pipe both elastic and inelastic particles behave the same, giving rise to an helicoidal motion throughout the outlet, however the delay between the two swarms of particles is already established and, while elastic particles are already close to the device outlet, the inelastic ones are still crossing the region between swirl chamber and pipe.

At first glance it might seem that, provided the collisions are inelastic enough, the correct residence time of the API particles is achieved. However, there are some concerns:

- In Appendix C we have demonstrated that, if the presence of the boundary layer at the carrier surface is neglected, then only a small underestimation of the characteristic residence time of the fine API particles is obtained. This legitimates us to release the API particles directly in the cup without the need to simulate the carrier particles. However, looking at the path the particles take through the device outlet, see Figure 11 (b), it is clear how they could spend most of their time inside the inhaler walls boundary layer as well. Thanks to the helicoidal motion inside the pipe, they could in principle cross the entire device without ever exit the boundary layer. As the viscous sub-layer is not simulated explicitly but heuristically though the wall functions, the drag force for the particles travelling in the mesh elements adjacent to the walls could be overestimated, i.e. the residence time underestimated. Panel (c) of Figure 11 shows the particles radial distance from the main inhaler axis, for a single time instant while the particles are crossing the pipe, as a function of y coordinate (the cup is located at y=0, the device outlet roughly at y=5 cm). The black continuous line represents the radial distance of the pipe wall (notice how the pipe is not a perfect cylinder having a slightly larger diameter at the outlet than at the swirl chamber intersection), the dotted line represents the full boundary layer of thickness $\delta$ previously estimated, finally the region between the continuous and dashed lines represent the viscous sub-layer which is modelled through the wall function. Indeed, regardless of the $e$ value, a significant amount of API particles is found in the viscous sub-layer where the drag forces are overestimated. Still, they do not travel permanently inside the viscous sub-layer, they rather enter and exit from it due the collisions with the inhaler walls. On one hand, if the permanence of the particles in the viscous sub-layer is short and given that, differently from the case of the detachment from the carrier surface, here the API particles are not at rest having their own inertia, the wrong drag force might be unable to alter significantly the particle trajectories. This might justify why the $d = 3\ \mu m$ peak, with $e$ in the range $0.1 \div 0.05$, overlaps with the measured data despite the incorrect drag force value in the viscous sub-layer region. The included diffusive forces due to the particle-eddy interaction can also play a significant role in this respect. On the other hand, the calculated residence time might match the measured one just due to a compensation effect: an overestimated drag force shortens the residence time, i.e. the emission peak should be shifted to the left compared to the correct one however, having chosen an $e$ value small enough, the peak is right-shifted matching the measured data.

- We released a very limited amount of API particles, in real products a much larger number of particles is present and the solid phase density, especially in the early stages of aerosolization, can be high enough to slow down the air flow. Thus, more realistic simulations require not only a larger number of particles but also a 4-ways coupling. Again, the good match between calculated and measured emission peaks, despite the reduced number of simulated particles could be due to a compensation effect. A simulation with the proper particle density and full coupling with the fluid might lead to a reduced air velocity and thus to longer residence times, compared to the present one. However, the selection of a larger $e$ value might still place the calculated peak in the right place.
- The specific surface area of fine API particles is very large and adhesion forces prevail over gravity, as a result such fines are usually found aggregated. Having no direct information neither on their aggregation state at rest on the carrier surface nor during their detachment, it is hard to estimate which is the best way to represent them at the beginning of the tracing. To avoid further hypothesis, we simply injected particles of different diameter to verify how strongly the residence time could be affected by the fact that the fines leave the carrier surface remaining aggregated in large “flakes” rather than in the form of individual particles. Results are shown in Figure 11 (d), the larger the particle/aggregate size the longer the residence time, a clear inertial effect. Also in this case, the fact that the $d = 3$ μ$m$, $e = 0.05$ peak matches the measurements might be both an indicator of the way particles behave during detachment from the carrier, or just the result of a compensation. On one hand it is possible to think that truly the API particles disaggregate and leave the carrier surface in the form of single particles with $d = 1 \div 5\ \mu m$. On the other hand, if the API remains in the form of aggregates with effective $d > 5\ \mu m$ the calculated residence time should be larger than the measured one, but the proper choice of $e$ (larger compared to 0.05) could still shift the emission peak to the left placing it over both the measure and the $d = 3$ μ$m$, $e = 0.05$ ones.

To conclude, despite the simplicity of the proposed approach and the many missing ingredients contributing to a more realistic description of the API fines behavior, an apparent good agreement between the calculated and measured residence times is found. However, based on the currently available experimental data, it is not possible to discriminate whether the agreement is true, and the strong approximations made really have a negligible impact on the residence time estimation, or it is due to the choice of a too small $e$ which compensates for the missing physics. Still the model is quite robust: having only one calibration parameter which allows a flexibility of no more than *10-12* ms on the residence time*,* i.e. less than 20% of the experimental value. For many of the inhaler engineering applications described in the last section of the paper the current model provides a good starting basis. However, if the aim is to better unravel the API disaggregation mechanism and its behaviour in the early moments of the device actuation, more experimental data on the state of aggregation of the API at the device outlet are needed. Only with such data in hands it makes sense to complicate further the API modelling.

## 3.4 Validation of the coupled CFD-DEM model

In this section we analyze the simulated emission of carrier particles and its dependence on the DEM model parameters. The results from the numerical model are compared with the reference experimental data summarized in Appendix A. As shown in Figure 9 (b) the average carrier particle diameter is comparable to the boundary layer thickness $\delta$, however it is much larger than the viscous sub-layer. Contrary to the API particles described in the previous section, the carrier particles center of mass always lays in the resolved region of the boundary layer, thus no significant error is introduced in the calculated drag and lift forces due to velocity overestimation. Another major difference is that, while the API fine particles follow the curling stream lines of the air flow, the carrier motion is dominated by inertial effects: the centrifugal force experienced by the particles while entrained in the swirling flow, squeezes them against the inhaler walls. Any geometrical feature can thus provide an obstacle for the particles to continue their helicoidal motion towards the outlet, this is for instance the case of the sharp step-edge at the connection between the swirl chamber and the pipe. The carrier particles trapped there can be further accelerated by the air flow enhancing the tangential component of their velocity $\overrightarrow{v^p}$. On the other hand, inelastic collisions can reduce the particles kinetic energy favoring the radial component of the drag force (dependent on the radial component of the slip velocity $\overrightarrow{v_{slip}}$) to move them towards the inhaler center, i.e. overcoming the edge obstacle and continuing their path to the pipe and outlet. This competition between the centrifugal force and radial drag is at the basis of particle classification in many industrial applications such as cyclone classifiers and jet-mills.

As described in Table 3, three are the parameters involved in the DEM model of particle-particle and particle-wall interaction. We have performed several simulations varying them one by one to assess their individual role in the resulting particle motion and residence time calculation. Having no details on the microscopic interactions among particles and between particles and walls, we set the same values for both the particle-particle and particle-wall parameters. The results are summarized in Figure 12 (a), (d) and (g) showing the total emitted dose mass as a function of time compared with the reference experiment (continuous black line). Panel (a) shows the effect of reducing the restitution coefficient: highly inelastic collisions remove too much kinetic energy from the carrier particles forcing them to remain in the close vicinity of the walls, where the air flow is weak, the experienced drag force is thus small enhancing their residence time. On the contrary, see panels (b) and (c), the much more chaotic distribution of carrier particles, typical of large $e$, allows them to populate regions of the fluid with larger velocity, speeding them up as well. The same behavior was discussed in the previous section for the API fine particles. Notice also how, modifying the restitution coefficient, both the position and the slope of the emitted mass curve change simultaneously, a value of $e = 0.5$ seems to give the best agreement with the measured data (black curve). Panel (d) shows the role played by the rolling friction parameter $\mu_r$: enhancing its value shortens significantly the carrier residence time and the overall carrier emission process. If, during particle-wall collisions, the rolling motion is hindered, particles can only slide against the walls. The large static and dynamic sliding friction force experienced by the particles ($\mu_s = 0.5$) lower their tangential velocity component reducing the "trapping" centrifugal force, thus they will leave easier and sooner the step-edge to enter the pipe resulting in a small residence time. This is particularly clear looking at panel (f): in the simulation with $\mu_r = 0.1$ most of the particles trapped at the step-edge have a larger tangential velocity which prevents them to escape, with $\mu_r = 0.8$ some particles lower considerably such value, being able to enter the pipe (the black arrow indicate them on the scatter plot). Any value of $\mu_r \geq 0.5$, together with $e = 0.5$, seems to grant the best adherence to the measured data. While, from a numerical perspective, the rolling friction coefficient is just a handle to enhance or reduce the effect of sliding friction, from a physical point of view it is reasonable to expect a high $\mu_r$, given the highly irregular shape of the carrier particles that certainly hinder their rotation upon collision. Finally, panels (g)-(i) confirm the dominant role played by $\mu_s$: if its value is too small particles remain trapped at the step-edge rotating at high speed, for values larger than $0.5$ the sliding friction slow them down allowing the radial drag component to free them. Above $0.5$ the dose emission profile is no longer affected by the $\mu_s$ value. Notice also how, varying $\mu_s$ it is not possible to compensate for a too small $e$ value. In conclusion the correct shape of the carrier emission profile can be obtained only with a very narrow range of parameters, namely for $e$ around $0.5$ and both $\mu_r$ and $\mu_s$ in the range $0.5 \div 0.8$. Such values are those able to reproduce the correct timing for particle trapping and releasing operated by the step-edge, which plays a major role in establishing the carrier dynamics.

Other authors found a similar behavior in similar inhaler geometries [21], also in their case DEM simulations revealed how the highest particle velocities and highest number of particle-wall collisions was achieved in the swirl chamber. They also reported about the significant sensitivity of probability distribution functions for particle velocities and collision energies upon $e$ and $\mu_s$.

As a further robustness test we verified how sensitive are the obtained results to variations in the particle-fluid interaction models. We tested two other drag models, besides the Di Felice one used so far, and we included also the Mei lift, results are summarized in Figure 13 (a). While the Di Felice and Gidaspow models are in perfect agreement, the Koch-Hill model seems to overestimate the drag force, leading to slightly smaller emptying time. As anticipated in the numerical methods section the Koch-Hill model has been tested and validated for $Re^p$ smaller than our typical ones. What is important to stress here is that, even a badly chosen drag model, can modify the simulated carrier emission profile by a very limited amount. The inclusion of the shear lift force gives also negligible effects, the role of the Magnus lift remains to be clarified, unfortunately such force it is not implemented in the current version of the DEM software. Other authors emphasized the importance of lift forces, especially for particles larger than $100\mu m$ [21], however they switched on both the shear and Magnus contributions simultaneously, it is thus not possible to discriminate between the two. The negligible effect we found for the shear lift suggests that the largest contribution comes from the Magnus one due to the high angular velocities achieved.

Forcing a 1-way coupling it is possible to evaluate which is the effect of the fluid slowdown on the carrier dynamics, according to many authors this is a poorly discussed issue in literature [21]. The 1-way carrier emission profile is shown in Figure 13 (a) and is much slower than both the 4-way coupling and the experimental ones. This might seem counterintuitive, the powder is in fact expected to slow down the fluid velocity, thus removing the 4-ways coupling should have resulted in a faster fluid, i.e. stronger drag forces and ultimately shorter residence time for the carrier particles. The non-trivial longer residence time is due again to the presence of the step-edge between swirl chamber and pipe: when many particles are trapped there the radial component of the fluid velocity is reduced by the particle-fluid momentum exchange. Smaller fluid velocity means lower centrifugal force and thus the possibility for the particles to escape the step-edge. In the 1-way coupled simulation, the fluid slowdown at the step-edge is suppressed, the particles experience a stronger centrifugal force preventing their escape. The fluid slowdown at the step edge is indicated by the black arrow in Figure 13 (b), where the difference between the fluid velocity fields in the 1- and 4-ways coupling is presented. Notice also how strong is the slowdown effect in the cup at the beginning of the simulation due to the high powder density, it can result in velocity differences up to 35 m/s. The slowdown gradually decreases in intensity as the aerosolization proceeds and the local powder density reduces, i.e. the fluid volume fraction increases again to 1. Clearly, when the device is almost empty the velocity difference goes to zero.

### 3.5 Considerations on aerosolization based on the CFD-DEM model

With both the fluid and particle dynamics models validated, and after developing some feeling on their sensitivity and most critical attributes, we can use them to investigate in details the dose aerosolization and the API disaggregation from the carrier. From the previously published data it is clear how the Nexthaler® inhalation performances are boosted by more than 10% (in terms of fine particle fraction) by the use of the breath actuated dose protector mechanism which releases the dose once the air swirl is already partially formed [73,74]. We can investigate which is the reason for such performance gain comparing carrier and fines dynamics simulated with and without the triggering of the dose protector. Figure 14 (a) shows the results of the lagrangian particle tracing for $d = 3\ \mu m$ particles: the blue curve is the same as in Figure 11 obtained releasing the particles at 45 ms, the yellow one is obtained releasing the powder at the beginning of the simulation, as if the absence of the dose protector was exposing the particles to the air flow since the very beginning of the air flow ramp. Although initially very weak the air flow is able to extract the particles from the inhaler much earlier than with the dose protector, the emission peak occurs roughly at 35 ms rather than 65 ms. Notice how the yellow peak sits over the small experimental peak anticipating the real fine emission: in the reference experimental work such peak is attributed to the aerosolization of some drug escaping the dose protector, its synchronization with the peak calculated in the absence of the dose protector is a further proof of the reliability of our calibrated model. Moreover, the possibility to reproduce both the peaks with a simple particle injection ignoring the slower carrier dynamics, suggests that the main emission peak is almost entirely originated by direct detachment, i.e. by the fluid drag and lift forces directly removing fine API particles from the carrier surface. The tail in the emission profile is then generated by carrier particle collisions and inertial effects which take place in a later stage and on a longer time scale. Panel (b) of the same figure shows the carrier mission profile from CFD-DEM simulations, also in this case the carrier emission starts slightly earlier without the dose protector, but both the emissions terminate almost at the same time. A sequence of snapshots of the carrier particle trajectories is given in Figure 15. Without the dose protector, the powder starts leaving the cup only after 10 ms from the flow ramp initiation. It takes then 10 ms more for the first particles to leave the swirl chamber entering the pipe, the same condition is realized after only 5 ms when the dose protector is present, panel (b). A top view of the simulations with the dose protector can be compared with the previously published experimental data showing the cup emptying, see Figure 15 (c), the agreement is remarkably good considering the adopted minimalistic DEM model. Finally, panel (d) shows a snapshot of the early stage entrainment of carrier particles in the absence of dose protector, notice how the carrier motion starts when the fluid structures are not yet fully developed, i.e. when the two initial vortexes have still to fuse together to form the final steady one. In the simulation with the dose protector, carrier motion is allowed only when the final central vortex is consolidated. Notice also how, in the simulations of both panel (a) and (b), the carrier particles in the proximity of the device outlet have a velocity of roughly 6 m/s, exactly as the value estimated processing the previously published PIV data (see Appendix A).

The properties of the fine emission by carrier collisions can be investigated analyzing the DEM particles trajectories: Figure 16 (a) and (b) show the collision number and the average collision velocities for both the cases with and without the dose protector. Due to the small number of carrier particles, particle-wall collisions are always more frequent than particle-particle ones, they are also more energetic, i.e. more effective in detaching the API fines. When the whole dose is accumulated in the cup or in the bottom of the swirl chamber the collisions are more abundant, as the aerosolization proceed and the particles get diluted the number of collisions decreases linearly. The yellow and blue bars above the plots mark the duration of the carrier emission, no significant differences are found in both the number of collisions and their energy during the carrier emission with or without the dose protector. The difference in inhalation performances cannot be attributed to the carrier collision detachment mechanism. One possibility is that the fines experience different drag and lift forces from the fluid, due to the different stage of formation of the vortex structure the carries is exposed to. This can be verified looking at the weighted average slip velocity $\langle\overrightarrow{v^f} - \overrightarrow{v^p}\rangle$, proportional to the average drag and lift forces, and the weighted average turbulent kinetic energy $\langle k\rangle$, proportional to the turbulence induced fluctuations in drag and lift forces:

$$\langle\overrightarrow{v^f} - \overrightarrow{v^p}\rangle = \frac{\int \left|\overrightarrow{v^f(\vec{r})} - \overrightarrow{v^p(\vec{r})}\right| [1 - n(\vec{r})]d\vec{r}}{\int[1 - n(\vec{r})]d\vec{r}}$$
$$\langle k\rangle = \frac{\int k(\vec{r})\,[1 - n(\vec{r})]d\vec{r}}{\int[1 - n(\vec{r})]d\vec{r}} \qquad (3)$$

the integrals being over the whole inhaler volume. Weighing the two averages with the volume fraction allows to retain the fields values only where the powder concentration is non-zero. Results are shown in panels (c) and (d) of Figure 16, also here there are no significant differences between the cases with and without the dose protector. In both cases, except the very early instants, the fines on the carrier surface are exposed to the same average drag and lift forces and to the same force fluctuations. The only remaining possibility, to justify the measured performance increment with the breath actuated dose protector, is that the fines detached from the carrier experience different fluid properties while traveling through the device, before reaching the outlet. Indeed, if we look at the turbulent kinetic energy density averaged over the whole inhaler volume $\hat{k}$, i.e. the same as in eq. (3) but without the weight over the carrier density, we notice a significant difference. This quantity is reported in Figure 16 (e), where the dashed lines represent the fines emission peak as per Figure 14 (a). Without the dose protector fines spend more time inside the inhaler, however, the average value of $\hat{k}$ felt is below 10 $m^2/s^2$, with a peak value of 15 $m^2/s^2$; the presence of the dose protector limits the fines residence time, however they feel an average turbulent energy above 20 $m^2/s^2$ with a maximum value of 23 $m^2/s^2$. In conclusion, based on these numerical evidences, it seems the increment of more than 10% in the inhalation performances, brought by using the breath actuated dose protector, is mainly due to a better chance of deagglomeration and dispersion for the already detached fines on their way to the device outlet.

## 4. Conclusions

In this paper we analyzed the aerosolization process for a model DPI in the NextHaler® comparing the experimental data available in literature with the results of CFD, lagrangian tracing and CFD-DEM simulations. We demonstrated how, for both the fluid dynamics and the solid phase behavior, it is possible to achieve a satisfactory quantitative agreement. We discussed which are the critical model parameters influencing the different aspects of the simulated aerosolization, from the CFD boundary conditions to the particle-fluid interaction empirical correlations, from the selection of the DEM contact parameters to the choice of insertion conditions for lagrangian tracing. We also discussed how to improve the model including currently missing elements. Concerning the description of the pure air flow behavior, the present RANS approach seems to catch correctly the main flow features and structures, it also work fine for the description of carrier particles inside the inhaler. Improvements can be done along the following lines:

- More complex turbulence models could be explored, especially those capable of including the anisotropic character of the large-scale perturbations (e.g. large eddy simulations) affecting the dispersion of fine API particles. Before considering them, given their higher computational cost and numerical stability issues, we felt it was necessary to start from the simpler, Reynolds-averaged, eddy-viscosity model here employed and discussed.
- The agreement between simulation and measurements, for both the steady state flow structure and the swirl formation during the initial transient, could be further challenged if more accurate experimental data were available. For instance, a higher sampling rate for the PIV at the device outlet could enable a comparison of the spectral analysis (Figure 7), allowing to verify the accuracy of the simulated time fluctuations caused by the vortex precession. A measure of the pressure drop occurring inside the device, close to the cup or in the lower part of the pipe could also help in validating the fluid behavior inside the device.
- What we noticed, in modeling the computational domain and applying the boundary conditions, is that the geometry downstream the device outlet has a significant impact on the development of the plume shape captured with PIV and fast camera movies, i.e. it influences the particle dispersion outside the device. On one hand this means the measurement apparatus geometry, even far from the device, must be accurately described in the future measurement campaigns (and a measure of both the inlet/outlet pressure drop and flow rate would be desirable). On the other hand, it means that, for the simulation of drug deposition in the lungs, the computational domain should include both the device and the patient's mouth for the proper description of the particle plume geometry and its behavior.

- Modeling explicitly the dose protector opening during the CFD and CFD-DEM simulation and the flow development is feasible, but it requires more experimental information on the dose protector motion and the opening time.

The carrier dynamics is sensitive to the choice of the three DEM parameters $e$, $\mu_r$ and $\mu_s$, we adjusted these values until the calculated emission curve matched the measured one. In this calibration process we noted how only a narrow region of the parameter space gives satisfactory agreement with the measured curve. In fact, varying each DEM parameter, results in multiple effects: the calculated emission curve shifts while change its slope and deforms. It is thus not possible to obtain two calculated curves overlapping with the measured one from two dissimilar sets of DEM parameters. In this first calibration attempt, to keep the model simple and minimal, we neglected the cohesion forces, further work must be done along these lines to extend the DEM model to high API content products. The carrier emission time is mainly determined by how much time the particles spend orbiting at the sharp edge between swirl chamber and pipe, trapped by the centrifugal force. Lagrangian tracing of fine API particles released in the proximity of the cup, supposed to be filled with carrier, is able to predict the measured emission peak by adjusting only one model parameter. Also on the solid phase modeling side, many improvements are possible if driven by further experimental data:

- The non-sphericity of the carrier particle and their large surface roughness could be included in the model by the proper corrections in the drag and lift empirical correlations. This requires however to set some geometry parameters that can be estimated only through a morphological characterization of the particles [75]. To limit the complexity of the problem this study should be conducted aerosolizing pure model carrier particles of controllable shape and sharp (almost mono-disperse) size distribution, without API or other fine excipients.

- Other shear lift models as well as the Magnus lift force should be implemented and tested.
- The study of API fines through hard-sphere lagrangian tracing is very promising, however it must be further investigated using more complex tools available through OpenFoam®. They allow a full particle-fluid coupling (4-way), the inclusion of lift forces and other types of particle-fluid interactions.
- Although it is very hard to know the state of agglomeration of the fine API or excipient components before the aerosolization, hypothesis for the lagrangian tracing initial condition can be made based on a time resolved particle size distribution measure at the outlet during aerosolization. Such kind of measure can be easily performed using, for instance, the Sympatech® INHALER 2000 module [76]. The initial agglomeration state of the traced fines could be tuned until a good agreement is found between their simulated aggregation state at the device outlet and the measured one.
- Finally, a lot of work could be done on the empirical correlations used in the drag and lift force calculations, correcting them for the presence of the inhaler walls in case of high solid phase density (small $n$ values). It is in fact demonstrated by most of the works in literature that, during aerosolization, API and carrier particles spend most of their time in the vicinity of the inhaler walls.

In the last section of this work we used the calibrated model to investigate origin of the known significant improvement of inhalation performances, when the breath actuated dose protector mechanism is present, compared to the case when it is deactivated. From the numerical simulations it seems that the inhalation performance boost is due to the enhanced turbulent kinetic energy density, experienced by the already detached API particles during their residence in the inhaler. This leads to stronger fluctuations in drag and lift forces helping in disaggregating API agglomerates. Carrier collisions seems not to be affected significantly by the breath actuated mechanism. This first evidences must be of course corroborated by further experimental and modeling work. Indeed, ccorrelating the in-vitro inhalation performance to the product behaviour inside the device can be very useful in different contexts:

- In case formulation changes or device geometry variants give rise to a significant difference in the measured fine particle fraction, the reason could be investigated by looking at the fluid and powder behaviour through the CFD-DEM model. As previously shown variations in the particle-particle and particle-wall collision statistics and energetics can be spotted as well as changes in the fluid structures or in their turbulence state. This kind of simulations can take less than 24 hours to be performed, many of them can run in parallel for each different experimental variant. They will indeed be much cheaper than repeating the in-vitro inhalation experiments in a facility for PIV or fast camera movies acquisition, and, once they are deemed reliable, the simulations convey a richer amount of information even in regions, or at time scales, not directly accessible to experiments. This approach has been used in the recent work by Bass et al. [77] and Tibbatts et al. [78].
- Optimization studies can be set up to find which device geometry leads to the best inhalation performances. This procedure can be made fully automatic using optimization software, able to modify some geometry parameters based on a pay-off functions calculated through the simulations, and mesh adaptation tools, which will recreate the CFD mesh to the new geometry.
- The same automatic optimization procedure can be applied while keeping the device geometry fixed and allowing the optimization algorithm to control the formulation parameters, e.g. carrier particle size and shape, powder poly-dispersion etc. However, more interesting in this perspective, would be the device geometry optimization made on different products or formulations. It could be for instance applied to modify the geometry of a device, initially designed to deliver low product mass with low API content (such as small molecule delivery for COPD), making it suitable to handle higher powder masses per puff or higher API doses, i.e. higher adhesion/cohesion and low flowability blends (typically encountered in the delivery of bio-molecules such as anti-viral or insulin) [79,80].

Moreover:

- By simply modifying the time dependent boundary conditions, switching from artificial ramps to real spirometry profiles, the device aerosolization capability can be tested against different patient populations, e.g. with different degree of severity or paediatric. As a result, the device geometry can be tailored for specific patient population deeds in the spirit of a customized-therapy approach and a patient-centric product design.

- Combine the DPI aerosolization simulation, inside the device, with a lung deposition computational tool. CFD simulation of the upper airways and modelling of the lung deposition have progressed significantly in recent times [22,81–84], still in most cases the initial API disaggregation from the carrier and the further aerosolization are not accounted explicitly in the models. Coupling these two different pieces of the same physical phenomena might be helpful to better design the future DPI products against misuse, improving the patient compliance. It might also help in better understanding the effect of having a flow-rate dependent or independent fine particle fraction emission, or to clarify the role of extra-fine API particle content, which is certainly affecting not only the product behaviour in the lungs but also during the disaggregation/aerosolization in the device [85].

To conclude, this work could be certainly complemented and extended investigating the possibility to include, inside the CFD-DEM simulation, the disaggregation and dispersion of the API fines. This could be done by both:

- a multiscale approach, where the API is released as a continuum phase upon carrier collision and depending on the air turbulence state around the carrier particle, as recently suggested by some of the previously cited works;

- a direct CFD-DEM API particles simulation.

In the former approach the parametrization of the API release requires a lot of effort in understanding the behaviour of a single carrier particle covered with fines, their interaction with the surrounding turbulent fluid and their collisions. Moreover, treating the released fines as a continuum phase (Euler-Euler approach) it is not possible to account for further fragmentation and disaggregation of the release API lumps, as well as their possible size segregation or stiction to the inhaler walls. The direct CFD-DEM simulation, on the contrary will be feasible only after implementing the proper numerical framework to take care of the previously illustrated issues in handling a large number of particles (billions) having high speed and large poly-dispersion (1-5 µm compared to the 100-300 µm of the carrier particles). Notice also how in the direct simulation approach high-performance computing resources will be necessary to afford the large number of particles involved.

## Appendix A: Processing of the fast camera movies

The fast camera movies of both the particle plume side-view, at the device outlet, and the top view of the cup have been analysed with the standard image processing tools present in the commercial software Wolfram Mathematica 12®. Each photogram has been cropped and processed separately, an example of the sequence of different mathematical operations applied is shown in Figure A.1 (a). In the original grey-scale images each pixel intensity $I$ is represented by a number ranging from 0 to 255 (from black to withe) or, renormalizing, from 0.0 to 1.0. The binarization operation consists in setting the value of each pixel either to 0 or 1 depending whether $I$ is larger or smaller than a certain threshold value $I_{tr}$. After binarization the image is composed by a large amount of *0 pixels*, representing the background, and a number of *1 pixels* clustered together, representing the carrier particles. Cluster analysis algorithms enable to identify, number and analyse the clusters, measuring for instance their number and position in each photogram as well as their perimeter and area. Noise and spurious features are filtered out of the images by selecting only those clusters whose morphology satisfies certain conditions. The list of applied conditions is shown in Table A.1.

Two values of $I_{tr}$ have been tested: the smaller one is more suitable to highlight carrier particles over the dark background of the API fines plume, but it is not able to capture blurred particles over the white background, being thus less effective far away in time from the fine emission peak; the larger one, on the contrary, captures easily the particles over the withe background but loses all carrier particles hidden inside the dark plumes of the fine emission peak. The different behaviour just described is clearly appreciable comparing panels (a) and (b) of figure Figure A.1. The particle diameter is estimated knowing that each pixel is 32.68 μm long, according to the spatial scale presented in the original movies by Pasquali et al. [39]. The average particle diameter of each photogram is plotted in figure Figure A.1 (c) for both $I_{tr}$ values, it is a surface equivalent particle diameter, i.e. the diameter of a circle having the same area of the measured 2D particle projection. The average dimeter of the carrier employed in the experimental study, as well as in the simulations, is around 240 μm and is represented in panel (c) through the horizontal dashed line. A quick comparison reveals how the carrier equivalent diameter obtained by the image analysis is slightly overestimated, indeed the blending with the fine fraction might increase a bit the overall particle diameter, but more likely the overestimation is due to the approximation introduced in the binarization procedure. The particles emitted at the inhaler outlet are identified by differences between consecutive photograms, only those entering in the image frame are counted as new. The total emitted mass can be obtained by summing up in time the number of new coming particles multiplied by their average equivalent diameter and by their density $\rho_p$. The total emitted carrier mass is shown in Figure A.1 (d) for both $I_{tr}$ values, panels (e) and (f) offer a comparison with the intensity obscuration profiles adapted from the experimental reference works. Indeed, the image analysis method confirms that the emission of the first carrier particles already occurs few moments after the dose protector triggering, when also most of the fine component is emitted (highest intensity obscuration peaks in panels (e) and (f)). However, the majority of the carrier is emitted later, in correspondence of the third peak marked with the black arrow. The fact that most of the emitted material is carrier, is confirmed by the difference between the total obscuration intensity and the fluorescence one only. Being only the fine fraction marked with the fluorescent additive, the region where the two curves of panel (f) differ the most, is the region where the unmarked carrier obscures significantly the laser light. Notice how only the first part of the total emitted mass estimated via the image analysis has been plotted in panel (d). Some time after the carrier emission is started a large number of particles remains trapped in the camber downstream the device, where the movie is recorded. Due to the crowding in the photograms it becomes impossible to distinguish between new emitted particles, following straight trajectories from the outlet as indicated with blue arrows in panel (a) of figure Figure A.2, and carrier particles recirculating back and forth in the collection chamber. Two kinds of circular motion can be noticed looking at the video, a vertical one sketched by green arrow and a horizontal one highlighted with the yellow arrow, leading to a systematic overestimation of the emitted mass. Still, in the first moments of the dose emission, where the image analysis can be trusted, the total emitted mass curves are in good agreement with the other experimental data available.

Driven by the need of estimating the shape of the carrier emission curve even at larger time, far from the main fine emission peak, we decided to apply the image analysis method above described also to the top-view movies of the cup emptying. A value of $I_{tr} = 0.25$ has been employed, with such choice whatever remains in the bottom of the device and cup is not detected, only those particles climbing up the main pipe are counted. After image binarization we defined a normalized intensity as the ratio between the pixel pertaining to the particles and the total amount of pixels composing each photogram, if a large amount of particle is detected the normalized intensity will be high. The plot of the normalized intensity as a function of time is plotted in Figure A.2 (b), it shows a bell-shaped feature easily fitted with a gaussian function (green line). The non-monotonic behaviour of the curve can be easily understood looking at three different photograms in its ascendant, maximum and descendant portions. The photograms are shown in panel (d) together with the particles detected by the image analysis algorithms: in the ascending part most of the carrier particles are still rotating in the swirl chamber, trapped by the sharp edge connecting it to the pipe, here the look a unique fuzzy cloud and the image analysis algorithm is not able to count them; once the particles enter the pipe traveling to the outlet they are detected, rising the normalized intensity; finally as the device get empty no more particles reach the outlet and the normalized intensity drops back to zero. Although this normalized optical intensity has no direct connection to the emitted carrier mass, it gives a precise measure of the duration of the overall carrier emission time. We simply renormalized the gaussian curve in such a way that its integral corresponds to the 10mg of total dose mass, in this way its cumulative distribution represent the total emitted mass in time. Such curve is shown in green in Figure A.2 (c) and shows a good agreement with the data obtained analysing the side-movie of the particle plume, witnessing the robustness of the information obtained processing the movies. The data in Figure A.2 (c) will be used in the paper to validate the coupled CFD-DEM model.

Those carrier particles initially emitted in the empty collection chamber are easy to follow through subsequent frames and no confusion can be made with other particles trajectories. We can use the particle positions, stored during image analysis, to measure the distance they travel between two consecutive photograms and, knowing the frame rate of the movies, we can estimate their velocity. An example is given in Figure A.3 (a) and (b), for the case $Q_{max} = 60$ l/min and $t_{ramp} = 0.3$ s, the characteristic velocity of the first emitted carrier particles is around $5 - 6$ m/s.

## Appendix B: CFD technical details

To test how acceptable is the assumption of treating our air flow as incompressible we performed a steady state calculation with *rhoSimpleFoam*, a compressible flow solver, and compared the pressure drop and the velocity profiles with those coming from *pisoFoam*, our incompressible, time dependent solver of choice for pure CFD calculations. Mesh and initial conditions were the same used for the other calculations of the paper, the flow rate imposed at the boundary was $Q_{max} = 60$ l/m. No significant variations have been found in the calculated inlet/outlet pressure drop as well as in the average velocity and in the 2D velocity maps at outlet. We also verified how large are the temperature and density variations now that such quantities are explicitly solved and left free to change in the inhaler inner volume. Results are presented in Figure B.1 (a) and (b). The only significant variations are in the density distribution close to the cup and in the central part of the swirl chamber, here the relative difference can reach 5% once the steady state is established. Temperature variations are completely negligible. Panel (c) and (d) show the air velocity maps for the compressible and incompressible cases respectively. While all the fluid structures remain the same, significant variations of the velocity magnitude are visible in the central part of the swirl chamber and in the central part of the pipe. The core of the air swirl has a larger velocity in the compressible case. However, it must be stressed that this is just the picture once the fluid steady state is reached, by that time, most of the particles have been already emitted from the inhaler. Notice also that panel (d) represents a single time instant of a time dependent fluctuating field, while panel (c) is the outcome of a steady state calculation. Moreover, looking at the carrier particle trajectories, they pass through the inhaler adhering to the walls, far from the swirl core, thus remaining in the regions where the discrepancies in the velocity maps, between compressible and incompressible calculations, are negligible. To conclude: the air incompressibility assumption does not alter qualitatively the multi-phase flows and the carrier aerosolization; variations in the air flow velocity are minimal and maximized only once the steady state is reached; in principle the latter should not affect too much the carrier particle behavior; however only a coupled, compressible simulation could demonstrate this for certain.

As already mentioned in the methods section the experimental data have been collected applying a controlled flow ramp, ideally, also in the simulations we should drive the flow by imposing the same flow rate ramp at the outlet. Unfortunately, the applied flow rate as a function of time has not been measured. we thus started assuming a linear increase of the flow rate up to $Q_{max}$. With such assumption, the induced pressure drop profiles differ significantly in shape compare to the measured ones, i.e. those reproduced in Figure B.2 (b). As a second attempt we shaped the flow rate ramps assuming proportionality to the fluid velocity increase in the PIV data. With simple non-linear fits we came to the following expressions for the flow rate $Q(t)$ applied to the inlets as a function of time:

$$Q(t) = \frac{Q_{max}}{2}\begin{cases} \tanh\left(2.92\, t/t_{ramp}\right) + \tanh\left(7.68\, t/t_{ramp}\right) & for\ t_{ramp} = 0.3\ s \\ \tanh\left(3.33\, t/t_{ramp}\right) + \tanh\left(16.08\, t/t_{ramp}\right) & for\ t_{ramp} = 0.7\ s \\ \tanh\left(3.39\, t/t_{ramp}\right) + \tanh\left(40.93\, t/t_{ramp}\right) & for\ t_{ramp} = 1.2\ s \end{cases} \quad (A1)$$

Some of these profiles are shown in Figure B.2 (a) and have been applied as boundary condition to the red outlet patch sketched in Figure 1 (b). A zero-gradient condition is applied for both the pressure at the outlet and the velocity field at the inlet patches. All the other boundary conditions remain the same, as per Table 1. Figure B.2 (b) shows how the steady state values of the pressure drops (continuous colored lines) are in good agreement with the experimental data (colored markers), this confirms that the simulated geometry (the test rig one) has the same aerodynamics resistance of the real Nexthaler®. The adherence of numerical results with the experimental pressure profiles also in the initial ramp, are a proof of the good choice of the imposed flow rate functions (A1), the only exception is represented by the $Q_{max} = 60$ l/min case where some mismatch is visible. A closer look to the $\Delta p(t)$ profile for such case, in Figure B.2 (b) and (c), reveals a potential inconsistency with the experimental data. According to the measures by Pasquali at al. [39], for such flow condition the dose protector is expected to open at $45$ ms, i.e. a $\Delta p$ of 2 kPa should be reached at that time instant. In the simulation, however, $\Delta p$ is still $< 2$ kPa, this is clearly due to the uncertainty in the estimation of the flow rate ramps from velocity data. We thus decided to try driving the air flow by imposing the inlet/outlet pressure drop as a boundary condition, leaving the flow rate free to develop. The following analytical expression has been used to fit the measured pressure drops $\Delta p(t)$ and to impose them in the CFD simulation:

$$\Delta p(t) = p_0 \sum_{i=1}^{3} \left( \frac{2e^{-a_i \left(t/t_{ramp}\right)^2}}{1 + e^{-a_i \left(t/t_{ramp}\right)^2}} - 1 \right) \quad (A2)$$

The fit coefficients for the different flow rate ramps are given in Table B.1. The velocity field at outlet has been left free to change with a zero-gradient condition. At the inlet patches pressure has been set to the ambient one and the velocity field to a zero-gradient condition, as detailed in Table 1. Qualitatively, the fluid behavior inside and outside the inhaler is the same obtained applying flow rate driven boundary conditions. However, comparing the generated flow rate ramps in the pressure driven case with those imposed through eq. (A1) in the flow driven one, differences are found, as visible in Figure B.2 (c) and (d). In pressure driven mode the flow rate grows quicker, reaching $10$ l/min difference in certain instants of the initial transient, the discrepancy is maximum at $45$ ms (dashed black line), the moment in which, according to the experimental evidence, the pressure drop reaches 2 kPa and the dose protector should be released. Although they do not alter significantly the fluid behavior, the discrepancies highlighted in Figure B.2 might have an impact on the particle lift-up during aerosolization.

As a final verification step we repeated flow rate driven simulations using the yellow profile of Figure B.2 (d) rather than eq. (A1), only to find that no significant differences exist compared to the pressure driven case. To conclude: to be consistent with the dose protector release condition, the pressure driven mode has been selected and has been used for all the simulations presented in the paper.

## Appendix C: CFD-DEM coupling technical details

We start this appendix discussing the effect of the boundary layer on the drag and lift forces experienced by particles at rest at the bottom of the inhaler. As mentioned in the paper, the air velocity drops to zero approaching the walls of the inhaler, its flow passing gradually from turbulent to laminar. The thickness $\delta$ of the boundary layer, in which such transition occurs, depends on the velocity air reaches far from the walls (free stream velocity) $v_{\infty}$. At the beginning of the inhalation profile, when air is at rest or slowly moving, $\delta$ is very large, even several millimetres; as the air accelerates $\delta$ decreases significantly being 240 µm only once the steady state is reached at $Q_{max} = 60$ l/min (see Figure 10 and related discussion in section 3.3). In the boundary layer region the drag and lift forces exerted on the drug particles can be significantly attenuated due to their dependence on the slip velocity (relative velocity) between particle and fluid. This is particularly relevant when it comes to the API fines, whose diameter $d \ll \delta$. The drag and lift forces typically implemented in most of the CFD-DEM coupling codes or in the lagrangian tracers of CFD software do not account for the presence of the boundary layer slow down, this effect is automatically accounted only for large particles having $d \sim \delta$ if the boundary layer is explicitly simulated with a fine surface mesh. This is for instance the case of our simulated carrier particles whose size is comparable to the boundary layer which is partly resolved down to the viscous sub-layer, see Figure 9 (b). However, when $d$ is very small or if wall functions are used to describe the fluid behaviour at the walls, the drag and lift forces could be largely overestimated. The force calculation functions will in fact use a value for the fluid velocity which is the average over the whole mesh element (i.e. the whole boundary layer) or, at best, an interpolation between the mesh element vertexes.

Through a simple 1D model, described in Figure C.1 (a), we estimate the time necessary for fine API particles, starting at rest in the bottom of the inhaler, to reach the outlet 5 cm above with and without the inclusion of the velocity drop in the boundary layer region. The model only includes the drag or lift forces in the simplified formulation described in section 3.3, gravity is always negligible for the particle size considered. In the case where the boundary layer is neglected, the velocity used for drag and lift calculation is $v_{\infty}(t)$ which depends on the inhalation profile $Q_{max}(t)$ from equation (A1); when the presence of the boundary layer is included, equation (2) is used to estimate the velocity $v(x, y)$ as a function of the distance from the device bottom $y$ (while $x$ is set to be in the centre of the dose cup, equidistant from the two outlets) as long as $y < \delta$ and $v_\infty(t)$ for $y \geq \delta$. Examples of 1D particle trajectories moving from the cup to the device outlet under the effect of drag, with and without the inclusion of the boundary layer, are shown in Figure C.1 (b) and its inset zooming on the initial instants. A delay in the exit time can be easily estimated and plotted for both drag and lift forces, this quantity is plotted in panels (c) and (d) of the same figure for different particle diameter. The blue lines represent the results for the case in which the dose protector is not considered, i.e. the particles are subject to drag and lift since the start of inhalation, the effect of the inclusion of the boundary layer is negligible and the delay is always below 1 ms. However, if one looks at the particle residence time, the 1D models largely underestimate it compared to the experimental value of 63-65 ms (see Figure A.1 (e) and (f)). This is due to the fact that, in the realistic 3D swirling flow, both radial and tangential components of the air velocity drop to zero along the main axis of the inhaler, where the cup is located (see for instance Figure 6 (c)) such effect is not accounted for in the oversimplified 1D model, $v_\infty(t)$ is too large, drag and lift are too large and the particles too fast in leaving the inhaler. A correction factor of roughly $1/10$ can be applied to $v_\infty(t)$ in order to achieve a residence time of 65 ms for particles with $d$ in the range $1 \div 5\ \mu m$. Now that the particle lift-off is slower in time the delay introduced by the presence of the boundary layer is definitely more significant reaching almost 5 ms for both lift and drag as shown by the yellow lines in panels (c) and (d) of Figure C.1. If the presence of the dose protector is included, i.e. the particles are allowed to detach from the wall only at 45 ms, the delay is reduced between 1-2 ms, green lines of the same figure. As the flow is already partially developed and $v_\infty(45\ ms) \gg v_\infty(0\ ms)$ the particles accelerate quicker spending less time on the boundary layer, the effect of its presence is thus reduced. To conclude, the presence of the boundary layer increases the residence time of the fine API particles, the smaller the particle diameter the larger the effect. The delay in the exit time can be up to 5 ms over a total of 65, however, for our characteristic $d$ values and given the effect of the dose protector, in our specific application the delay is always below 1 ms. The error in disregarding the presence of the boundary layer in our CFD-DEM or lagrangian tracing simulations is thus negligible.

## Figures and captions

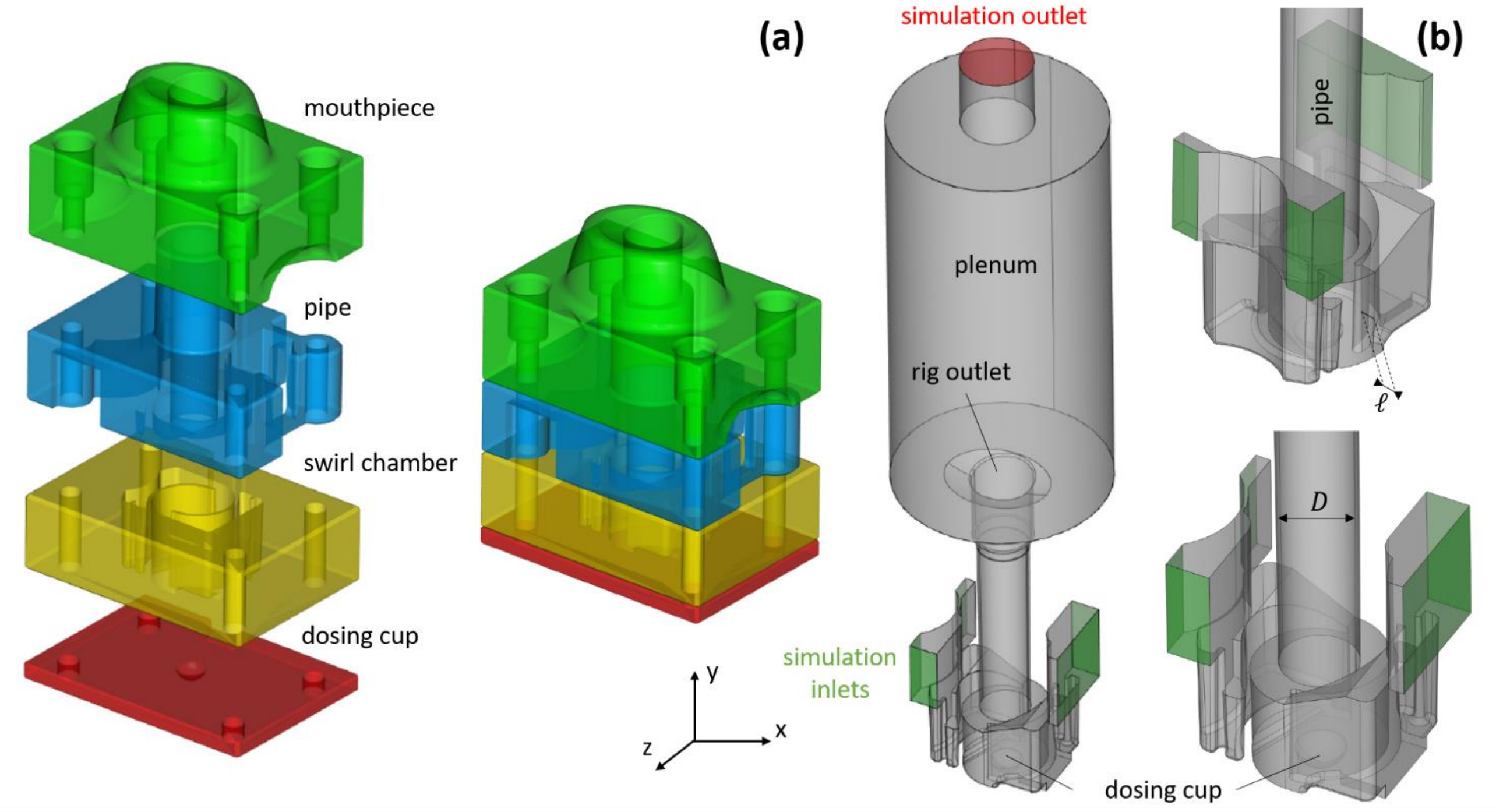

**Figure 1: Model Geometry**. (a) test rig assembled and exploded in its four sections. (b) Simulated fluid volume as the sum of the test rig internal volume and the discharge chamber of the experimental setup. The five squared flat patches marked with green color represent the inlets while the red circular patch represent the outlet.

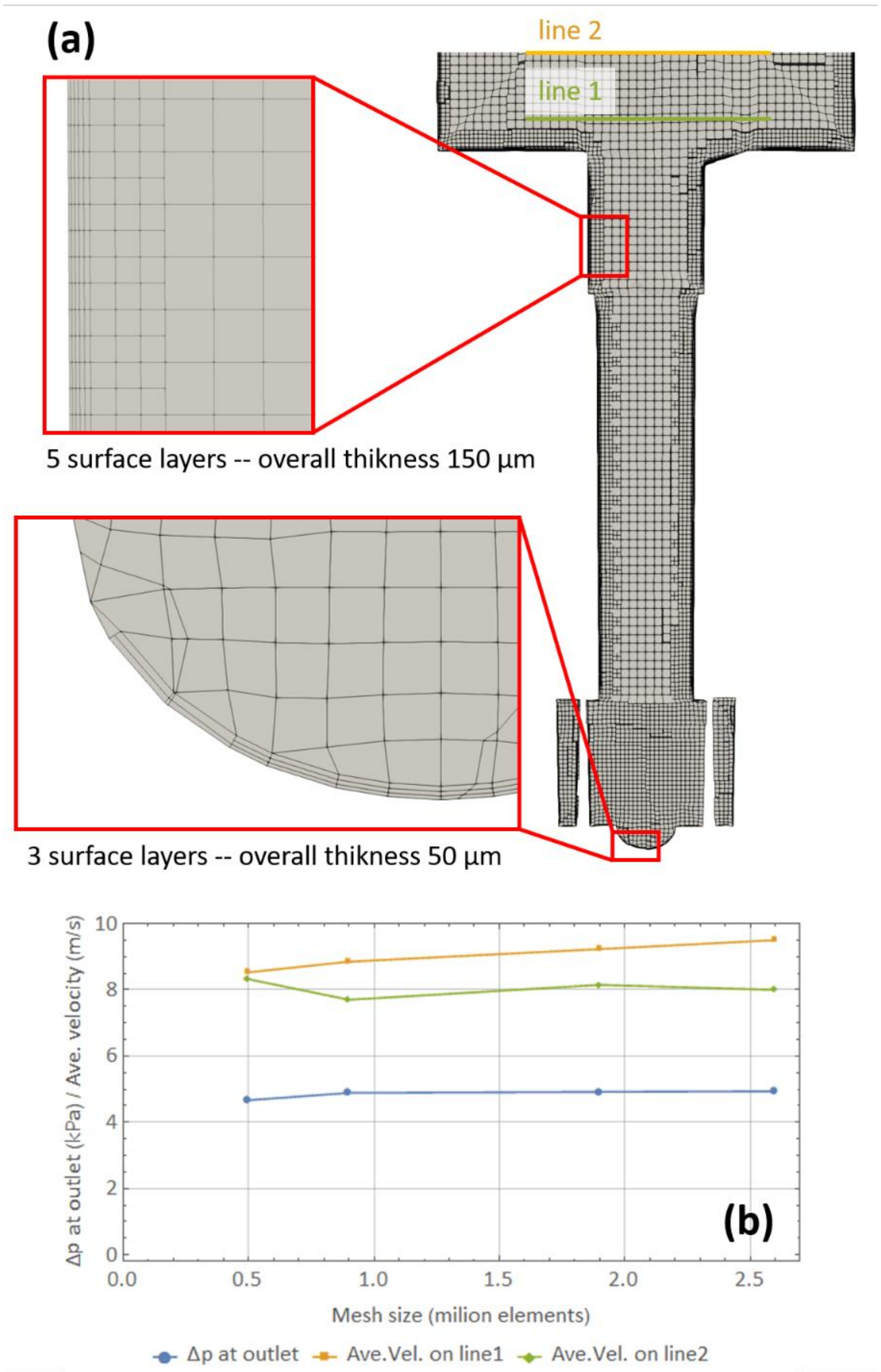

**Figure 2: Fluid volume discretization**. (a) mesh adopted for the CFD calculations, the insets show closeups on the finer surface elements along the walls. (b) Pressure drops between inlets and outlet and fluid velocity, mediated along the two lines shown in panel (a), as a function of the mesh accuracy.

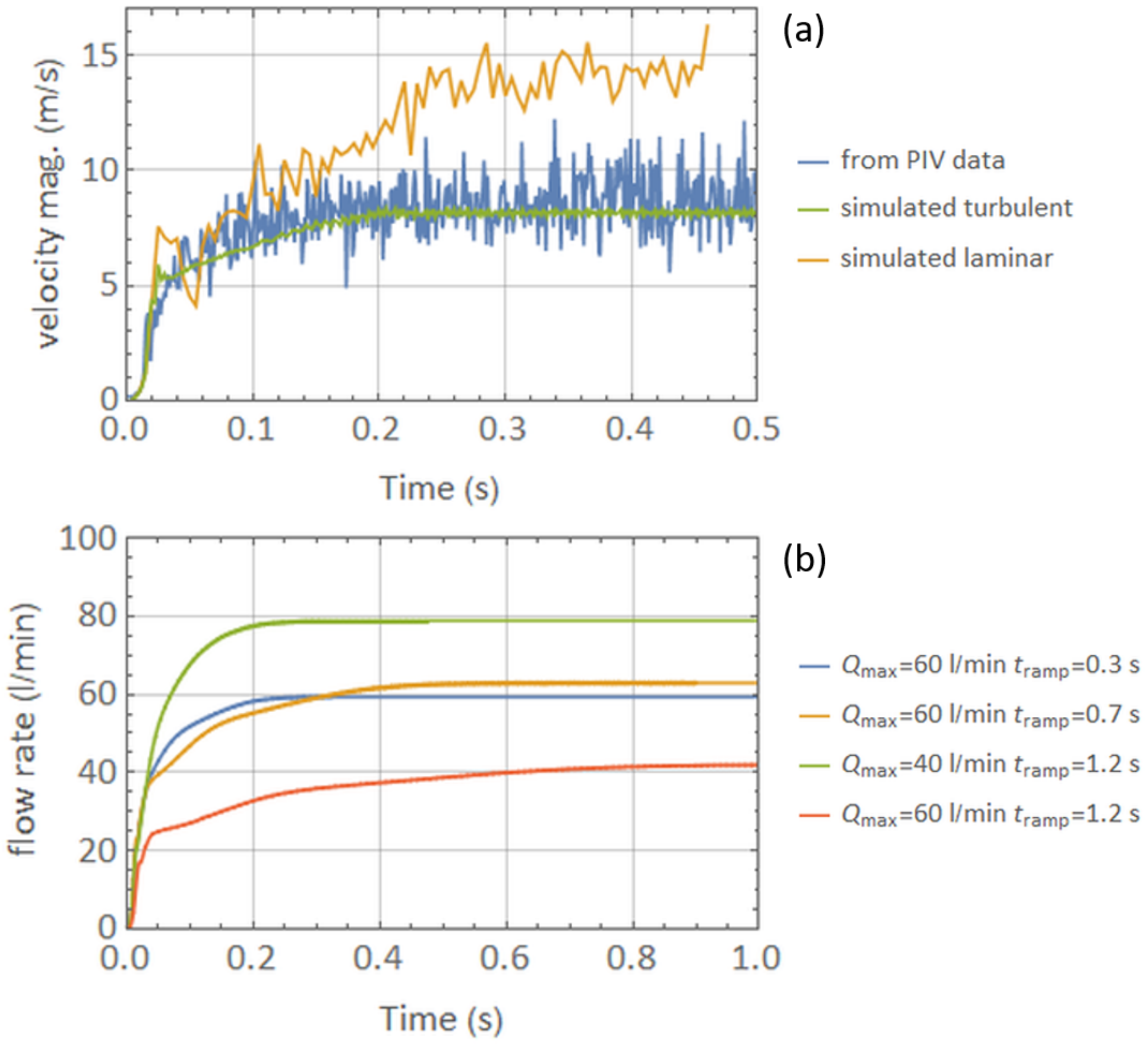

**Figure 3: Calibration data for the case $Q_{max} = 60$ l/min and $t_{ramp} = 0.3$ s.** (a) average air velocity computed at the inhaler outlet (on line 2 of Figure 2) from the PIV data (blue) and from laminar and turbulent simulations (yellow and green). (b) Computed flow rate profiles from CFD simulations at different inhalation conditions.

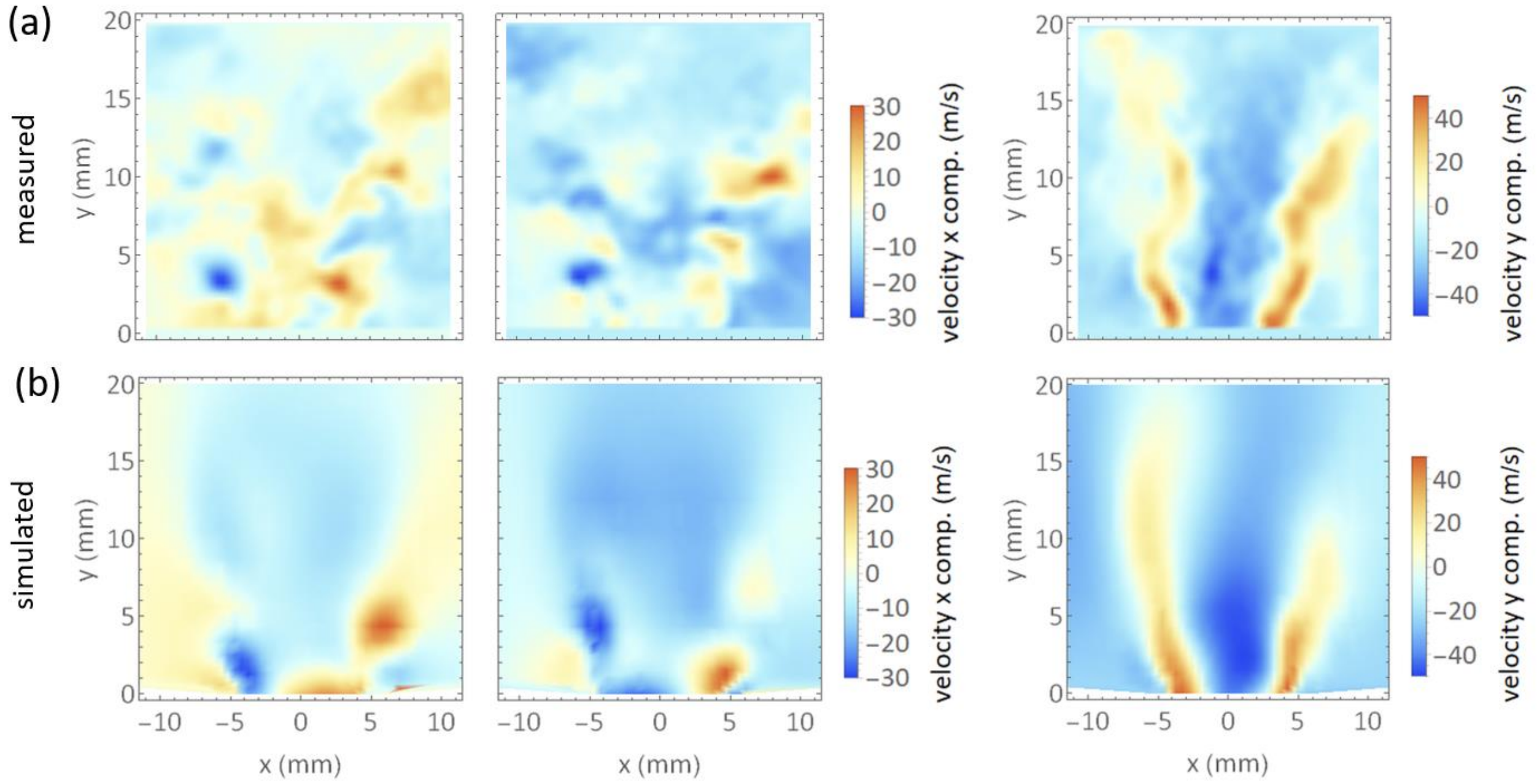

**Figure 4: Air velocity components at outlet**. (a) pressure drop across the device during aerosolization at different flow rate conditions. Markers represent measurements by Pasquali et al. while continuous lines of the same color refer to the simulated values in flow rate driving mode. (b) and (c) panels offer a comparison of the calculated velocity maps and the measured values through PIV experiments at the device outlet, they refer to the case $Q_{max} = 60$ l/min and $t_{ramp} = 0.3$ s. The central panels show the comparison for the x component of the air velocity while the right panels illustrate the velocity component along y direction.

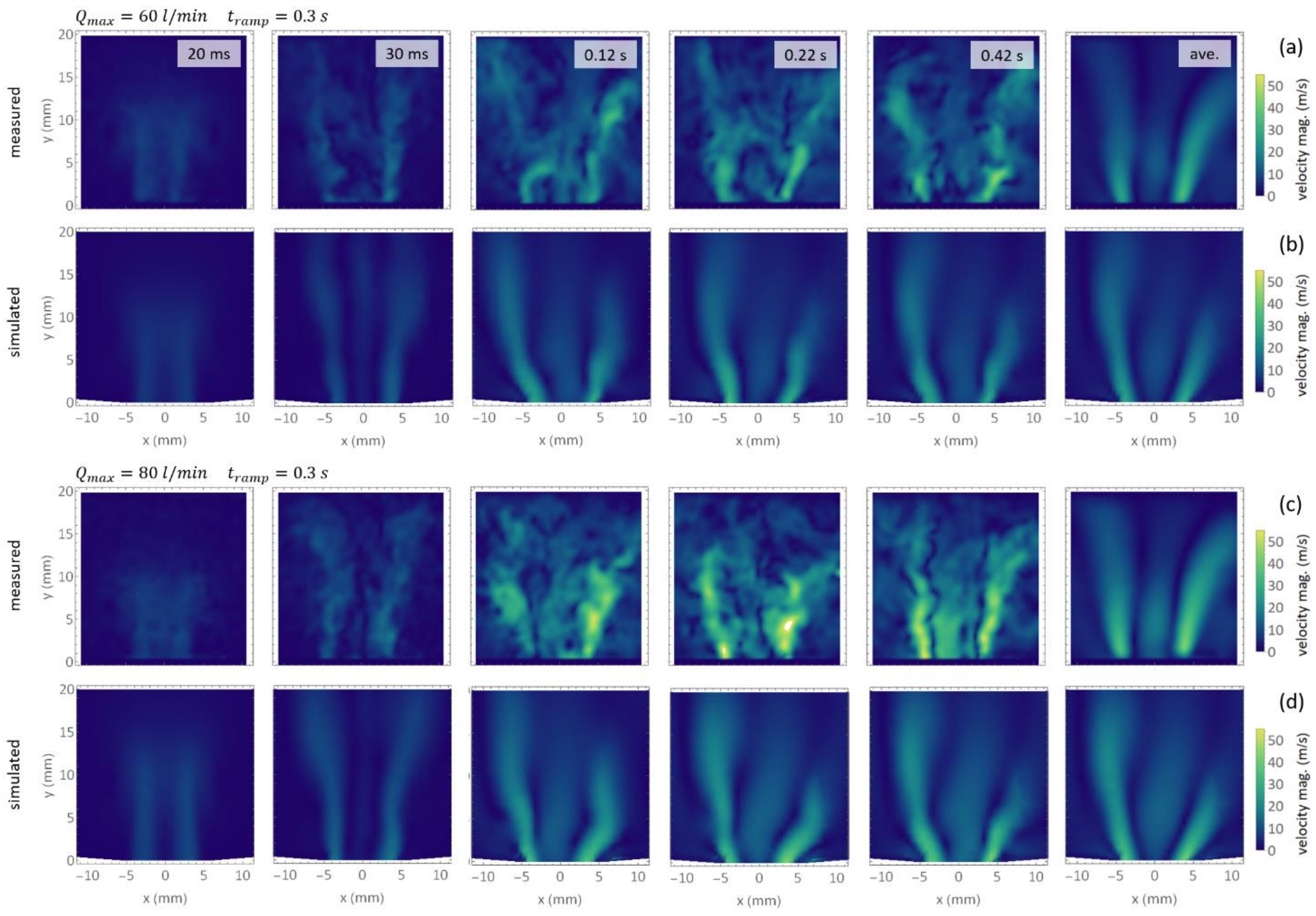

**Figure 5: Development of air plume at outlet**. Measured, (a) and (c), and simulated, (b) and (d), air velocity maps at the device outlet at different time instants during the aerosolization. The comparison is shown for two different flow rate conditions, indicated above the panels, at the same time instants indicated in the top row. The last panel of every raw shows an average of the velocity map taken over 20 ms once the steady state at the flow rate $Q_{max}$ is reached.

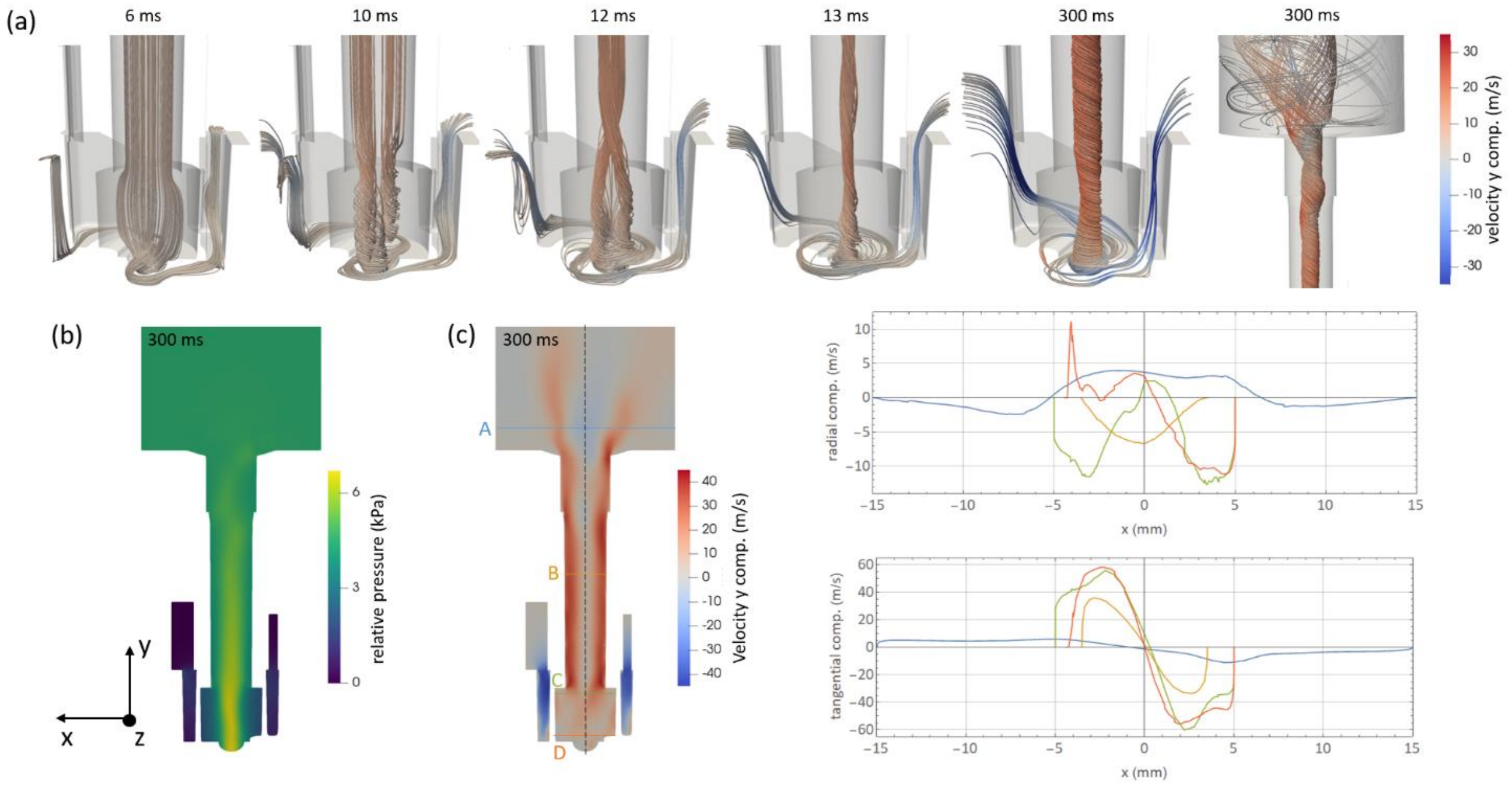

**Figure 6: Fluid behavior inside the inhaler for the case $Q_{max} = 60$ l/min and $t_{ramp} = 0.3$ s.** (a) streamlines showing time evolution of the swirling flow in the initial moments of the flow rate ramp and when the air vortex is fully developed. The streamlines are colored according to the y component of the fluid velocity. (b) pressure field in a cross section of the inhaler when the flow is fully developed. (c) axial (y-direction), tangential and radial components of the air velocity when the air vortex is fully developed. The black dashed line represents the main inhaler axis, the four colored lines with letters A-D highlight the direction along which radial and tangential velocities have been plotted.

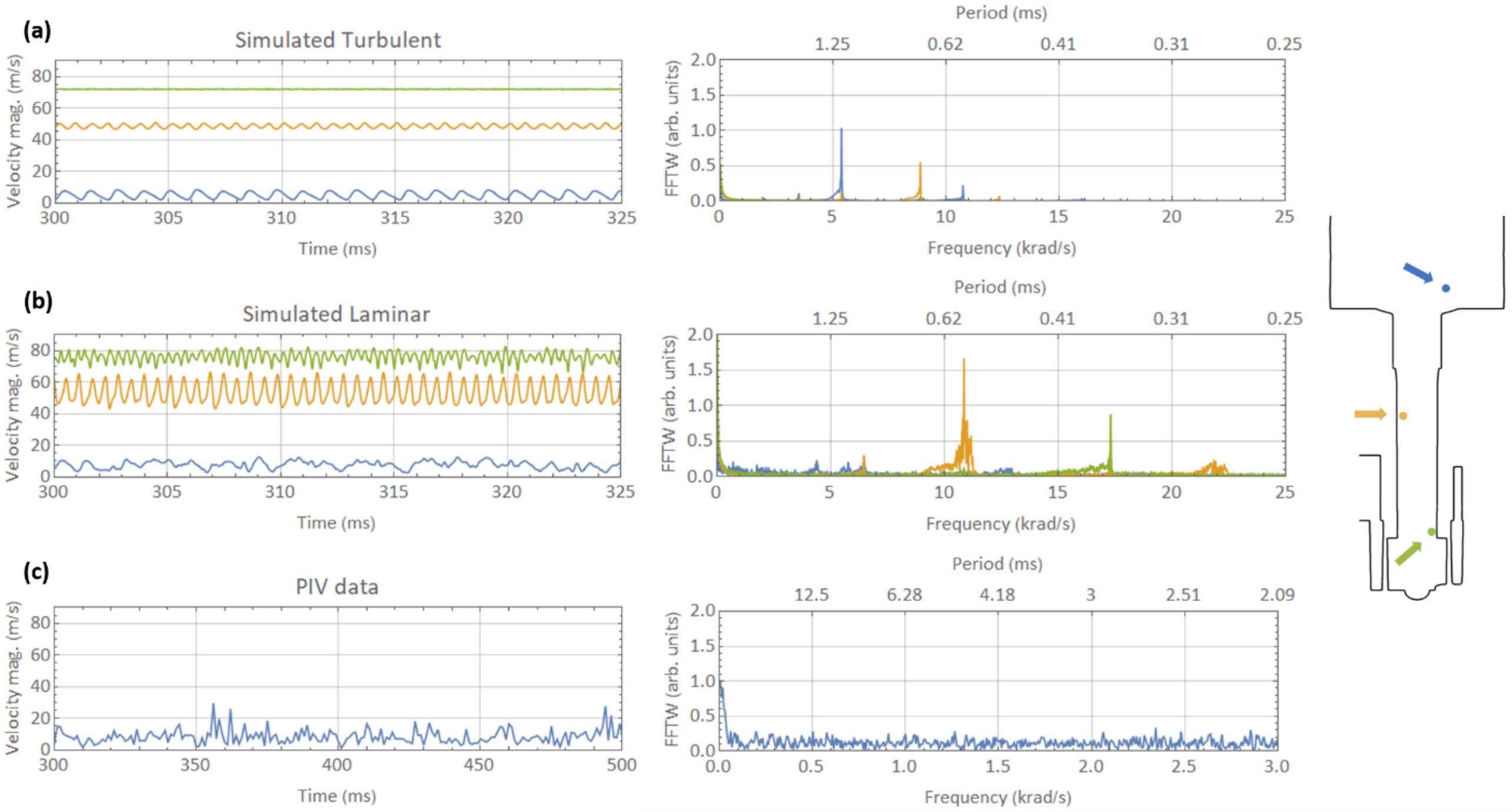

**Figure 7: Spectral analysis for the case $Q_{max} = 60$ l/min in the steady state condition**. Velocity as a function of time, recorded in the three points shown in the inset, and relative Fourier transform for (a) turbulent flow simulation, (b) laminar flow simulation, (c) extracted from the PIV data.

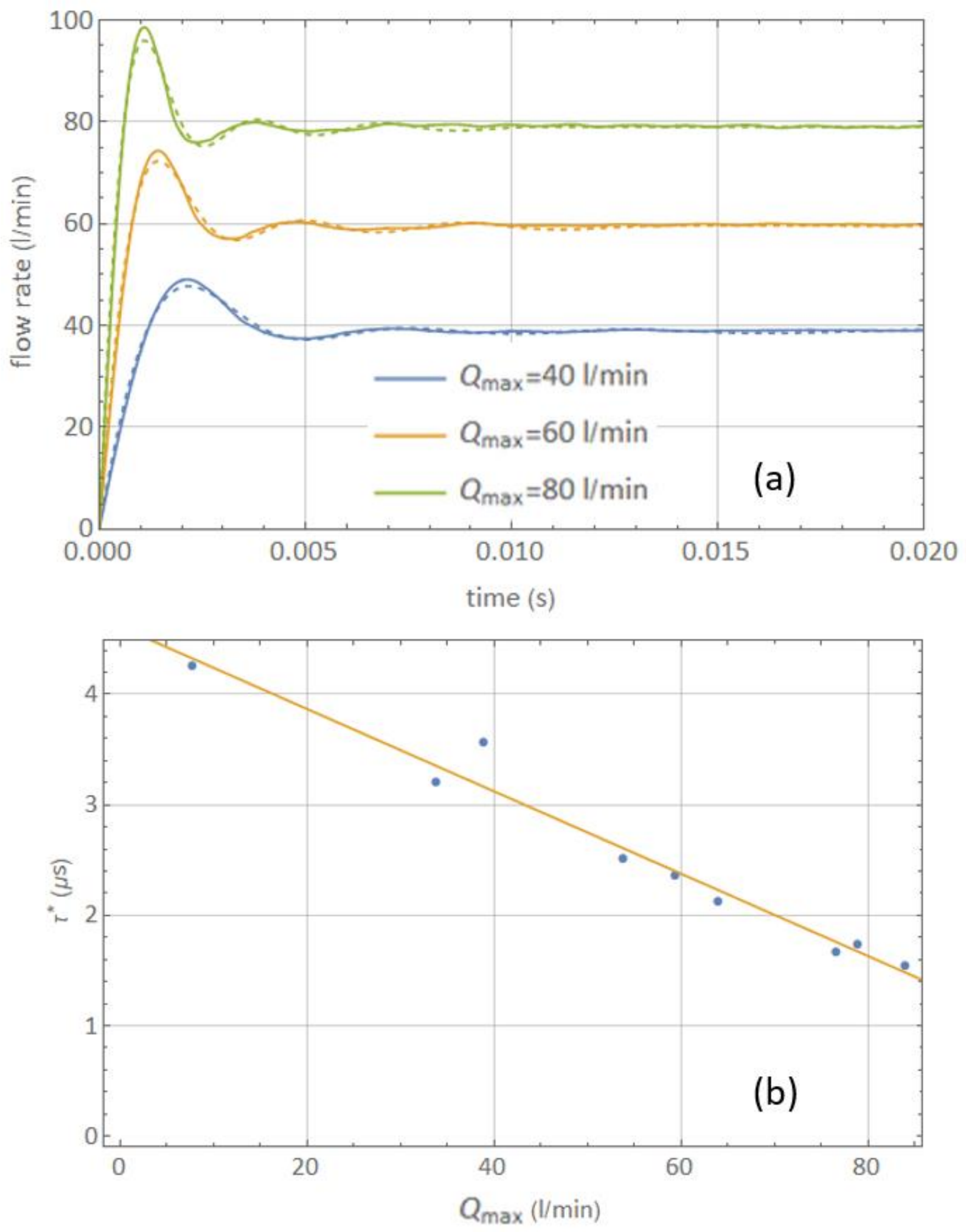

**Figure 8: Inhaler characteristic time**. (a) flow rate perturbation and relaxation in case of sudden switch-on of the inlet/outlet pressure drop for different $Q_{max}$, continuous lines are the results of numerical simulations while dashed lines are the fit with equation (1). (b) dots display the relaxation time for different $Q_{max}$ obtained fitting the CFD data like in panel (a), the continuous line is a linear fit.

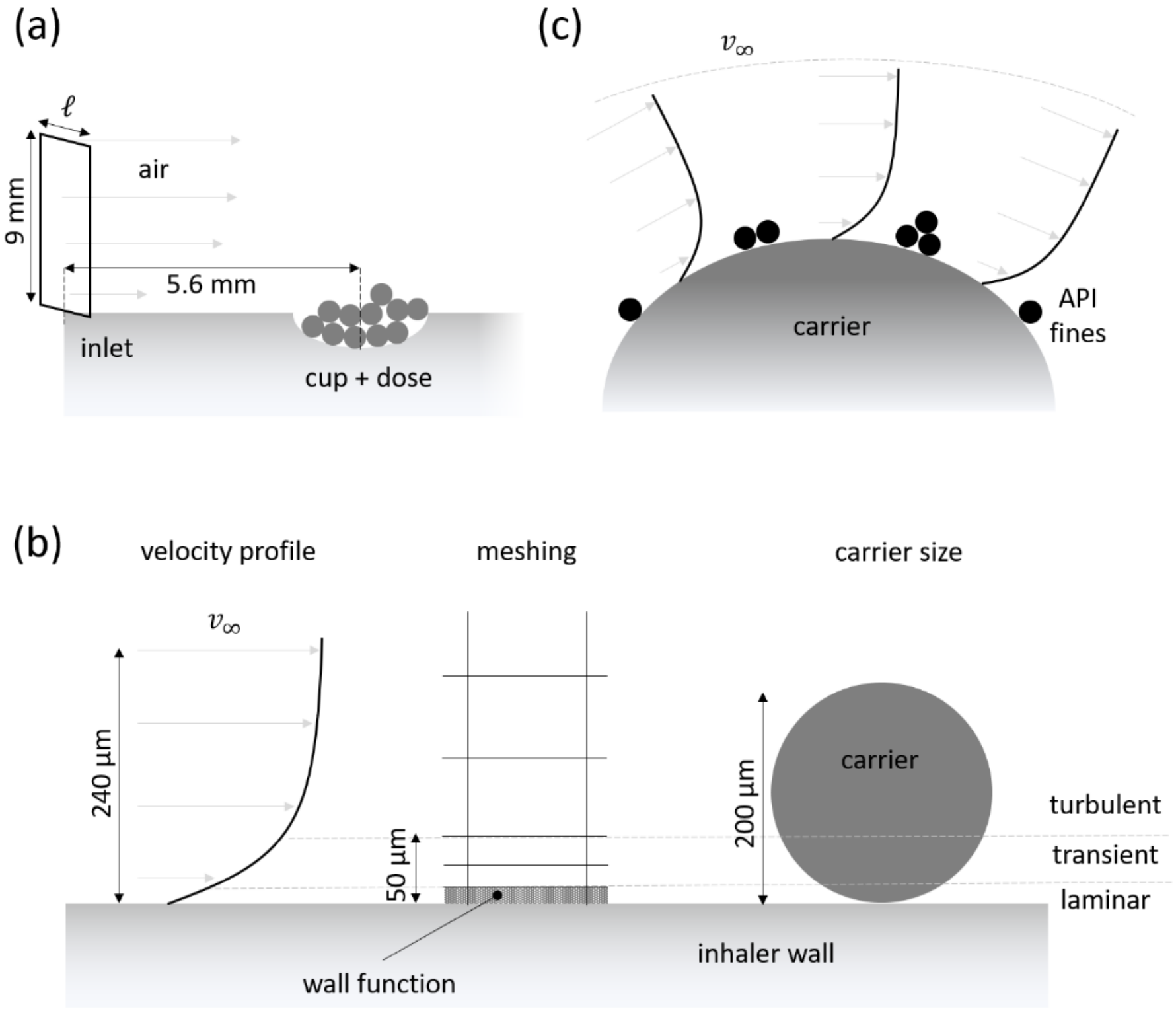

**Figure 9: Dose aerosolization and disaggregation**. (a) Sketch of the idealized 1D air flow from the inlet to the cup. (b) comparison of the spatial extension of the boundary layer $\delta$, the meshing scheme close to the walls and the carrier size. (c) boundary layer felt by API fine particles on top of a larger carrier particle, both the carrier curvature and the boundary layer are clearly not in scale compared to the size of the API spheres.

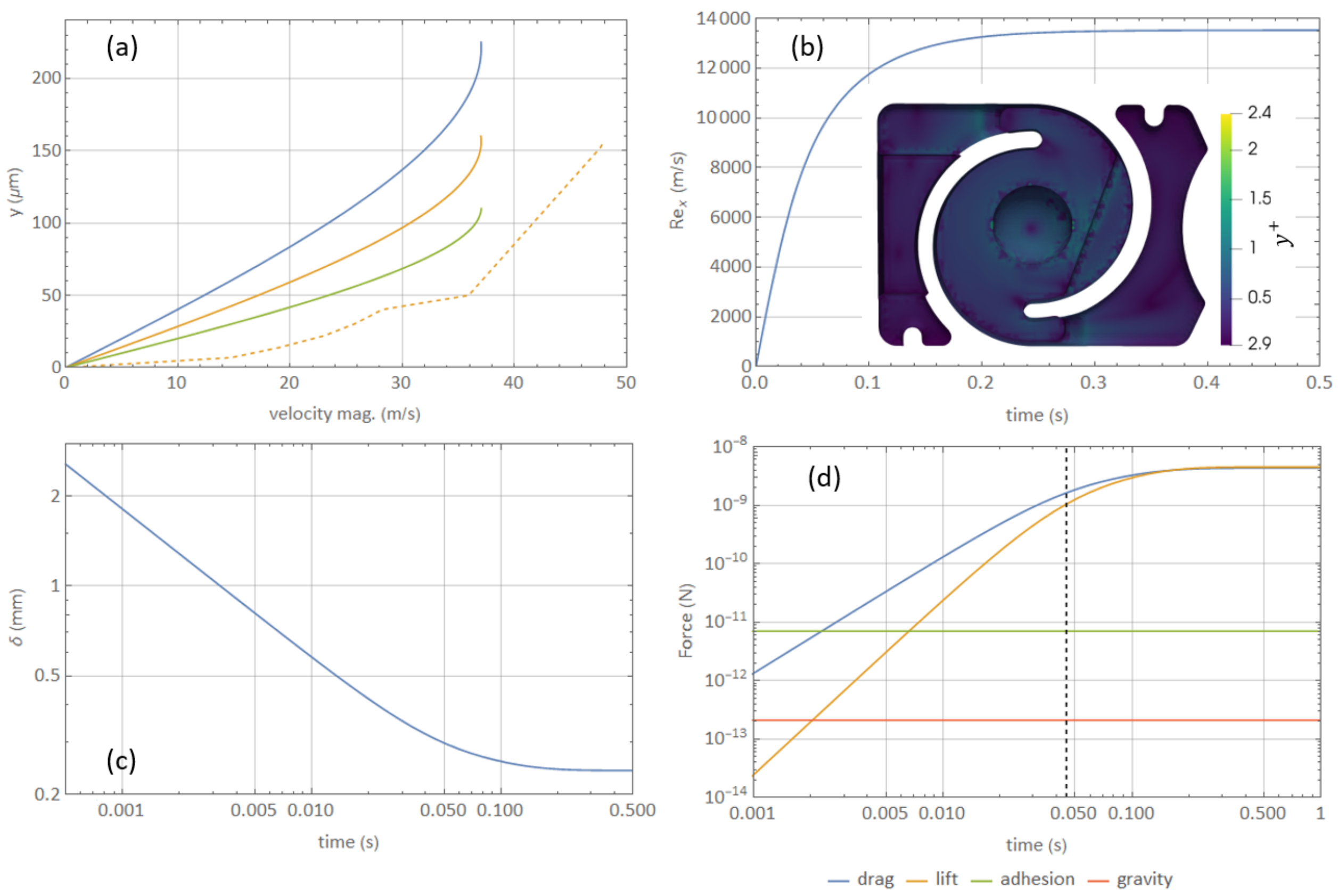

**Figure 10: Analysis of the boundary layer and particle-fluid interaction forces for the case $t_{ramp} = 0.3$ s and $Q_{max} = 60$ l/min.** (a) velocity profile along y estimated according to equation (2) at three different x values corresponding to the cup center (green), halfway between inlet and cup (yellow) and 1/4 of the distance between inlet and cup (blue). The yellow dashed line corresponds to the 3D flow from CFD simulations calculated halfway between inlet and cup. (b) and (c) temporal evolution of the local Reynolds number and boundary layer thickness at the cup center according to the 1D flat plate theory. (d) temporal evolution of drag and lift forces and comparison with gravitational and adhesion force estimate for $d = 3\ \mu m$ lactose particles. The black dashed line marks the dose protector shift instant.

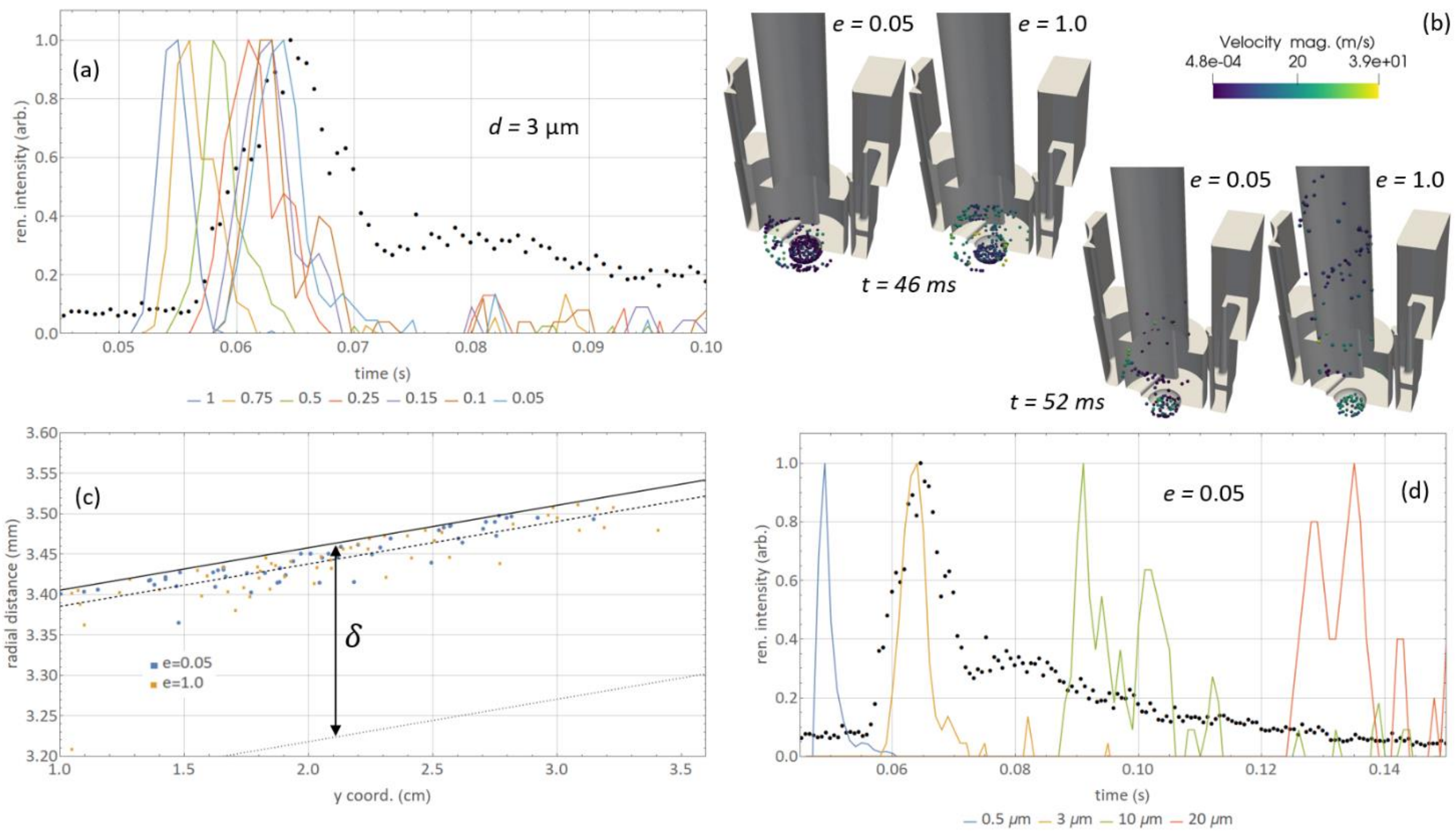

**Figure 11: Fine API emission**. (a) comparison of the simulated and measured fine emission, i.e. fine residence time. The black dots represent the reference experimental data, a normalized optical intensity proportional to the powder mass flow rate at the device outlet. The colored continuous lines represent calculated powder mass flow rate at the device outlet for particles with $d = 3\ \mu m$ and variable restitution coefficient $e$, like for the experimental data the curves have been normalized so that the main peak height is 1. (b) snapshots from the simulations of panel (a) taken at two consecutive time instants in the early stages of aerosolization. A comparison is shown between the two limiting cases of perfectly elastic and almost completely inelastic collisions. (c) particles radial distance from the main inhaler axis, oriented along the y direction, calculated for a selected time instant when particles are travelling along the inhaler pipe. The black line represents the radial distance of the pipe wall, the black dashed and dotted lines mark the whole boundary layer thickness and the viscous sub-layer one respectively. (d) same as panel (a) now keeping the restitution coefficient fixed to $0.05$ and varying the particle diameter.

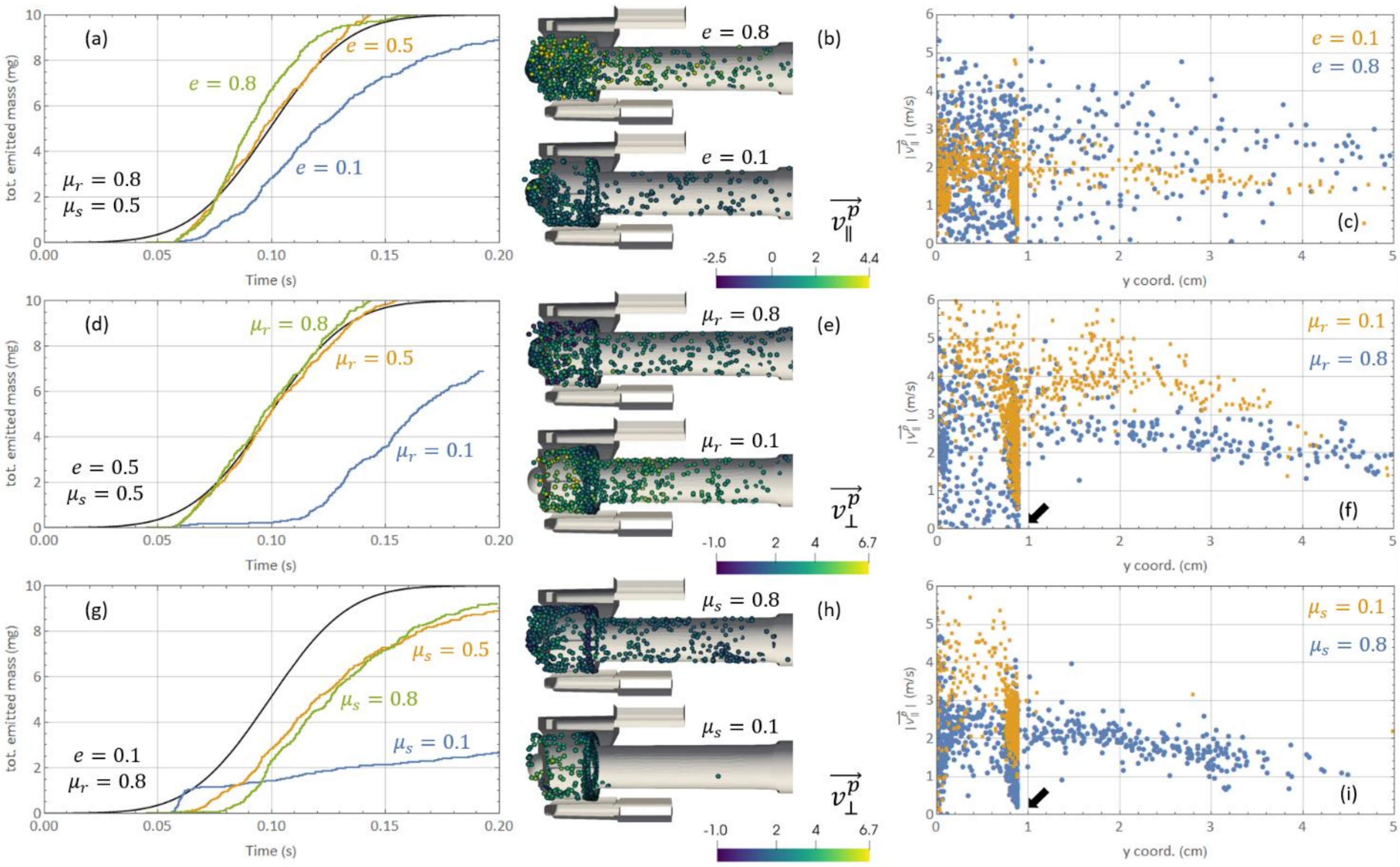

**Figure 12: Carrier emission**. (a), (d) and (g) show the total emitted dose (carrier) mass as a function of time for different DEM model parameters. The black line represents the carrier emission curve estimated from the reference experimental work data as illustrated in Appendix A. (b), (e) and (h) show the comparison of particle trajectories, at the same time instant, for the two extreme cases in the choice of restitution coefficient, rolling friction coefficient and sliding friction coefficient respectively. The color code represents the tangential or radial component of the particle velocity. (c), (f) and (i) illustrate the tangential component of the particle velocity at the same time instant as a function of the y coordinate (the dose cup at y = 0 cm, the device outlet at y= 5 cm).

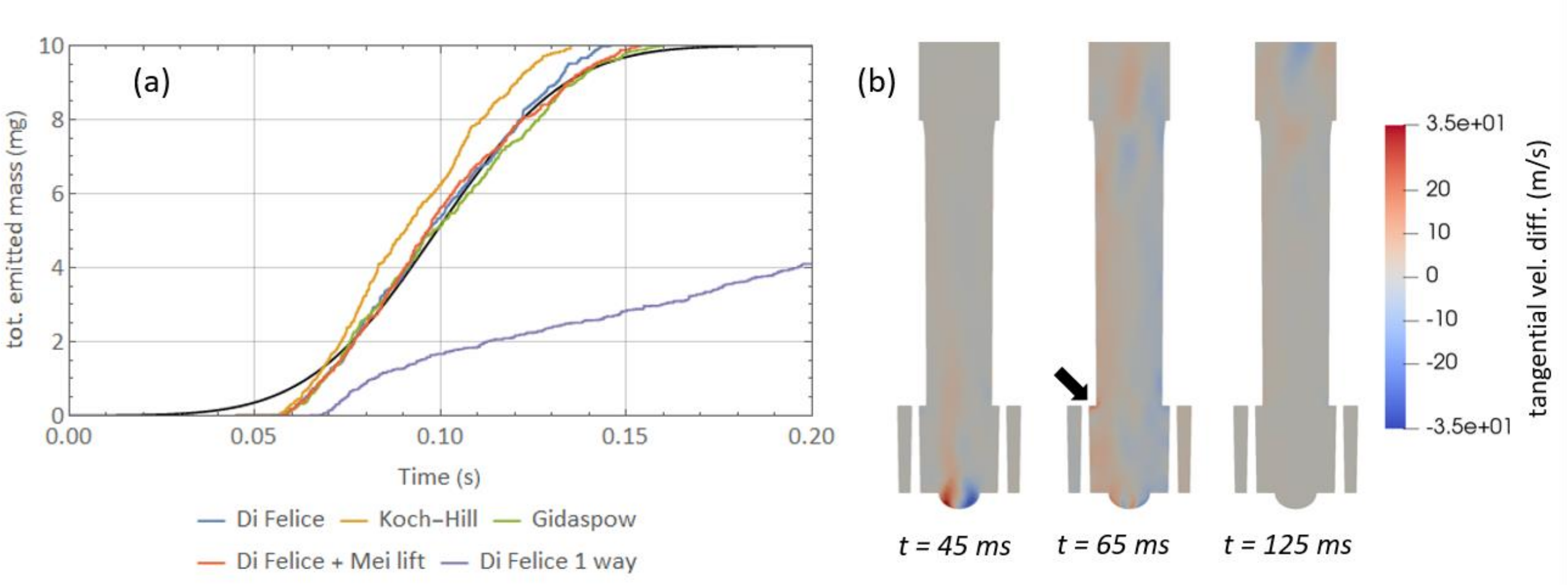

**Figure 13: 1-way vs. 4-ways coupling**. (a) total emitted dose mass as a function of time for different drag model, including shear lift force and forcing a 1-way coupling. (b) difference in the tangential velocity maps between 4-ways and 1-way coupling simulations at subsequent time instants. The following DEM parameters have been employed: $e = 0.5$, $\mu_r = \mu_s = 0.8$.

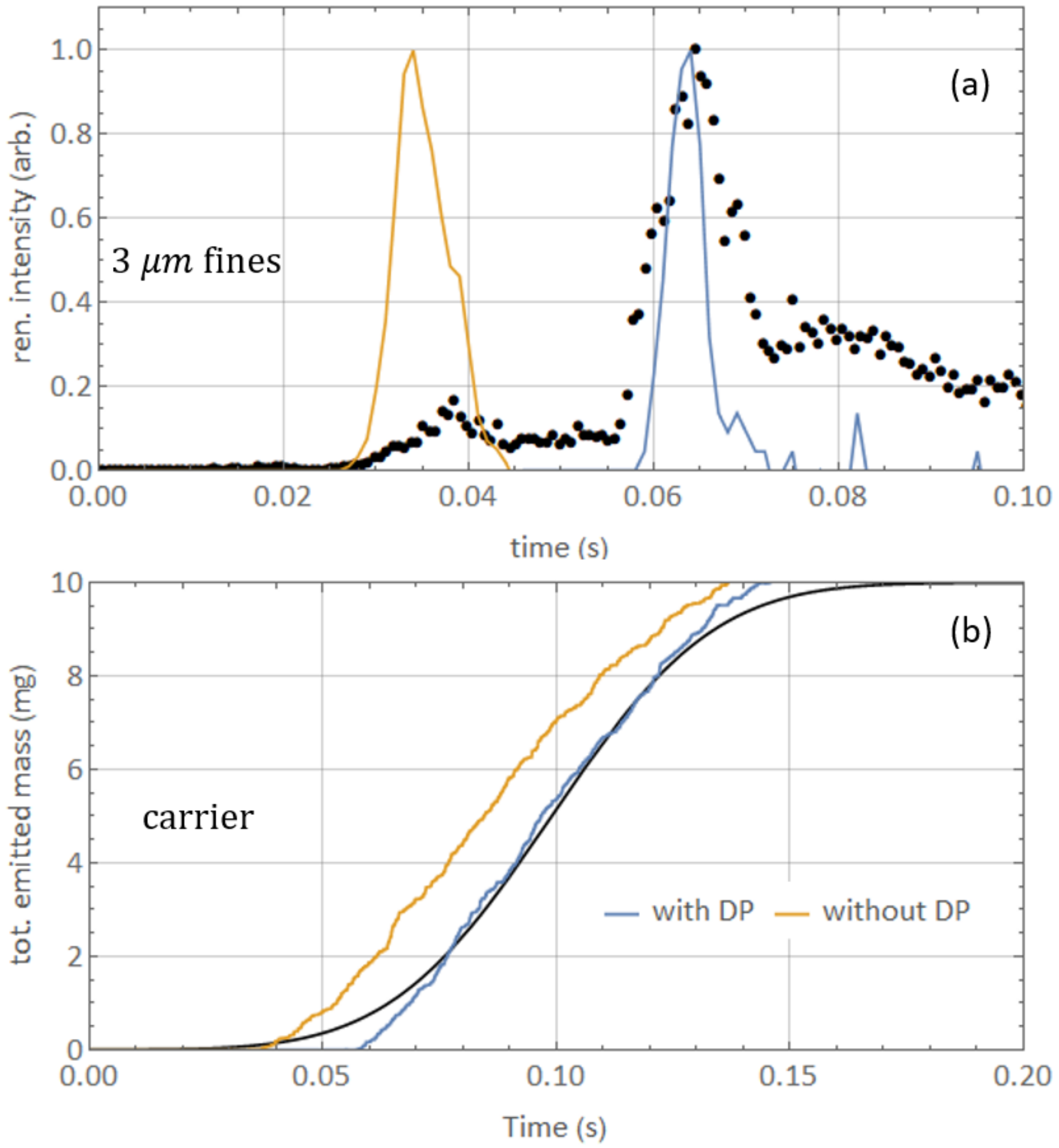

**Figure 14: Emission profiles comparison without (yellow) the presence of dose protector**. (a) fines emission profile as a function of time for 3 $\mu m$ particles through lagrangian tracing with $e = 0.05$, black dots represent the reference experimental data. (b) Carrier emission profile as a function of time from CFD-DEM coupling with $e = 0.5$, $\mu_r = \mu_s = 0.8$, black continuous line represents the experimental reference data.

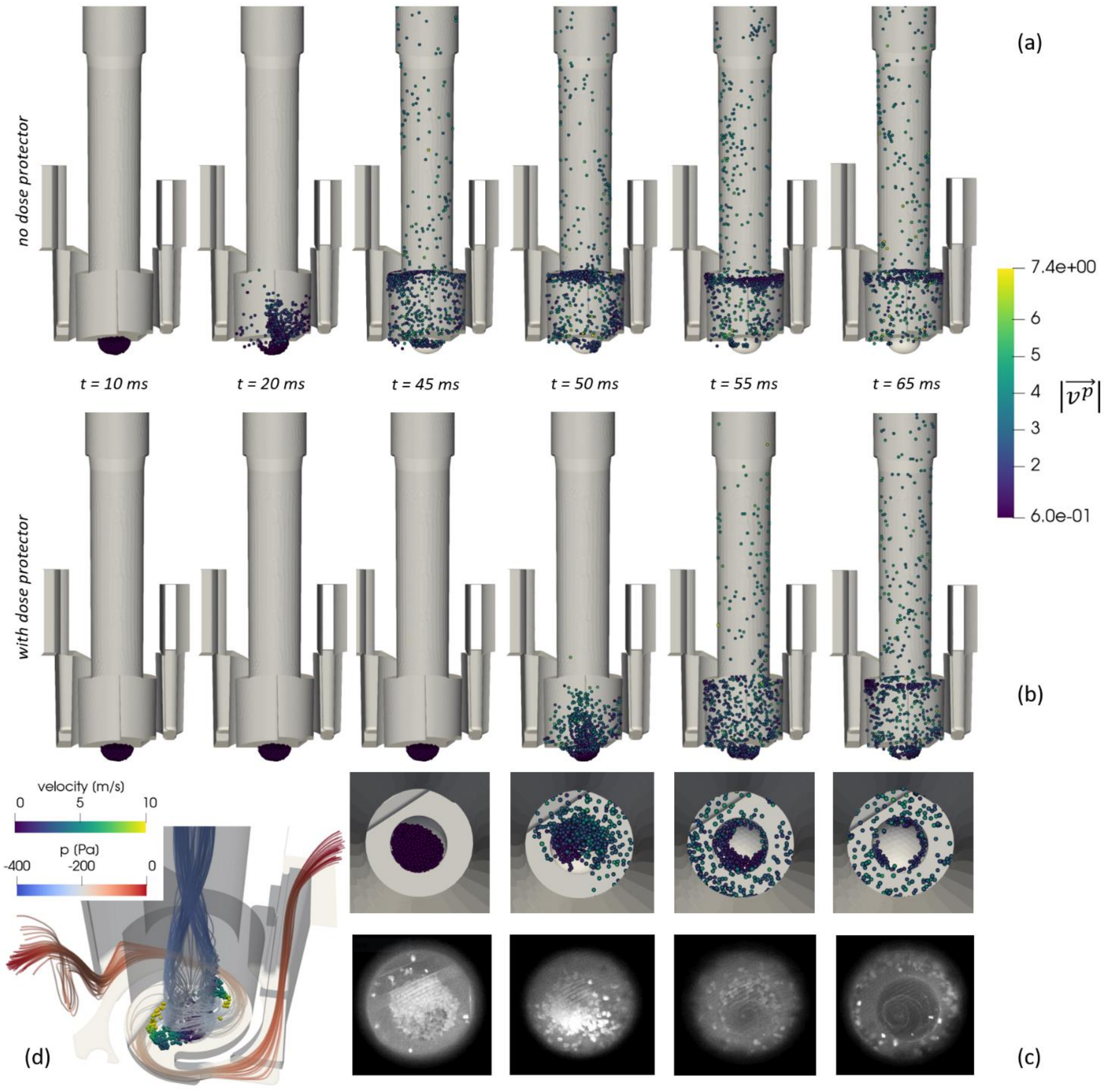

**Figure 15: Again on carrier emission**. (a) carrier particles trajectories at different time instants from the simulations without dese protector, the time of each snapshot is specified below the figure, the color scale represents the particle velocity magnitude. (b) same as panel (a) but for the simulation with the dose protector. (c) top view of the same snapshots of panel (b) compared with the experimental data presented in the reference experimental work. (d) snapshot at $t = 15\ ms$ from the simulation without dose protector, stream lines are colored according to the fluid pressure, particles are colored according to their velocity magnitude.

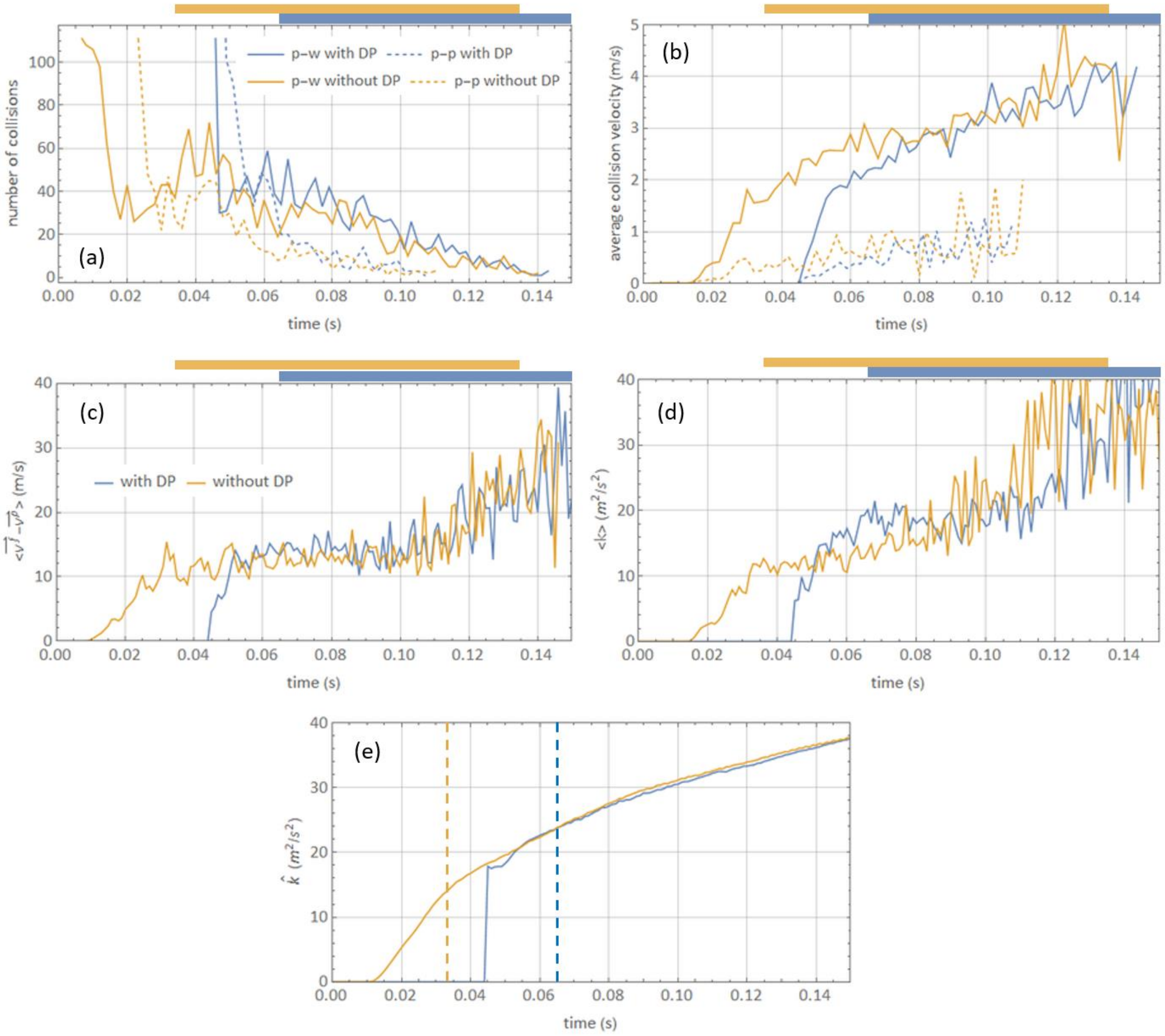

Figure 16: **Comparison of carrier and fine properties with and without the dose protector**. (a) number of particle wall (p-w) and particle-particle (p-p) collisions as function of time. (b) average collision velocity for p-w and p-p collisions as a function of time. (c) weighted slip velocity average for carrier particles as a function of time. (d) weighted turbulent kinetic energy density average for carrier particles as a function of time. (e) average turbulent kinetic energy density as a function of time.

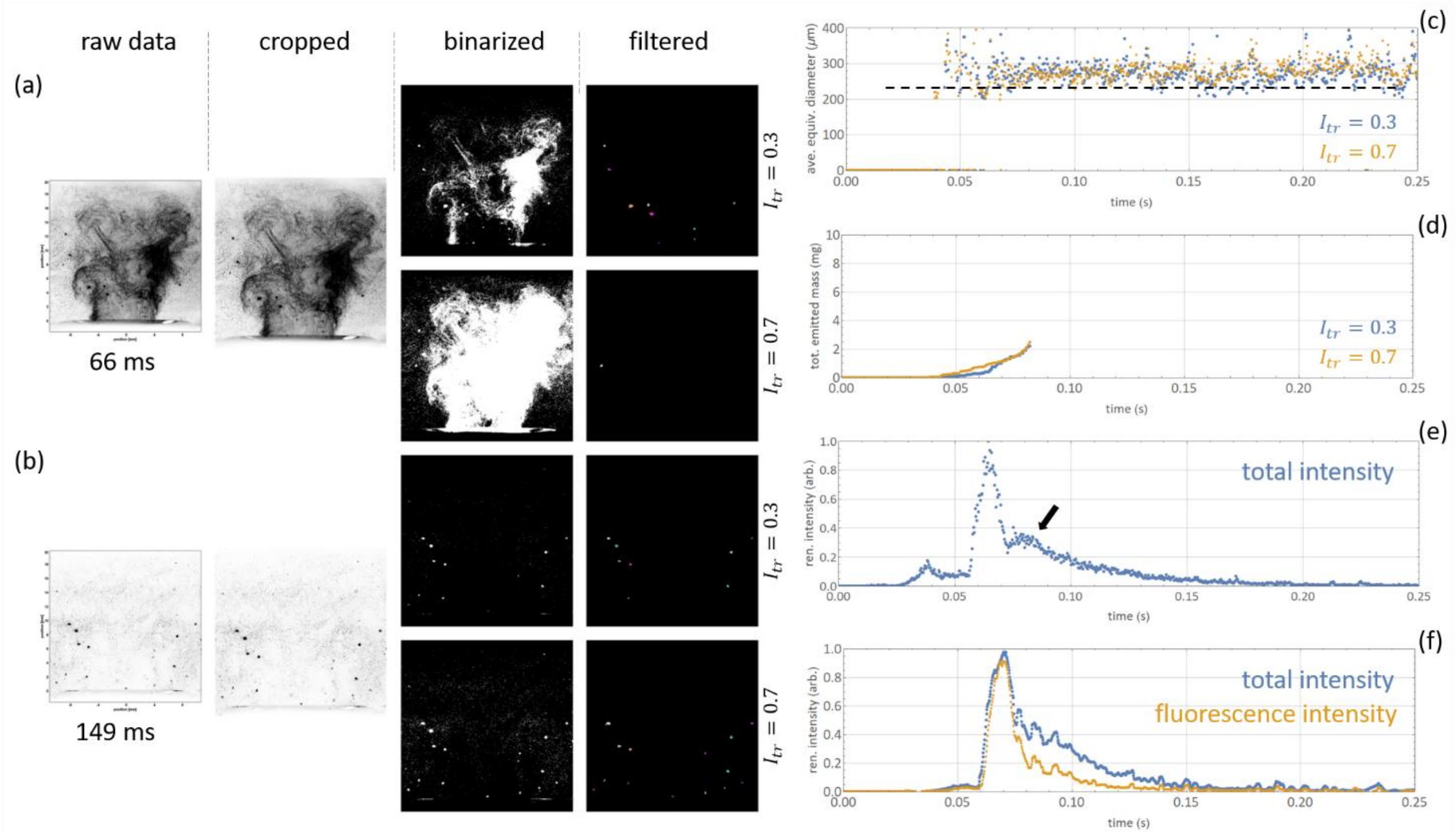

**Figure A.1: Image analysis of the side-view of the particle plume for the case $Q_{max} = 60$ l/min and $t_{ramp} = 0.3$ s**. Panels (a) and (b) illustrate the processing of two different photograms: the raw data image is first cropped and then binarized using two different $I_{tr}$ values. As a last step the image is filtered and only those clusters of pixels representing carrier particles are retained, each of them is coloured differently in the rightmost image. (c) average carrier particle equivalent diameter as measured by the image analysis algorithms. (d) carrier emitted mass as a function of time as estimated from the image analysis method. (e) and (f) intensity obscuration profiles adapted from the two experimental reference papers by Pasquali et al. and Merusi et al..

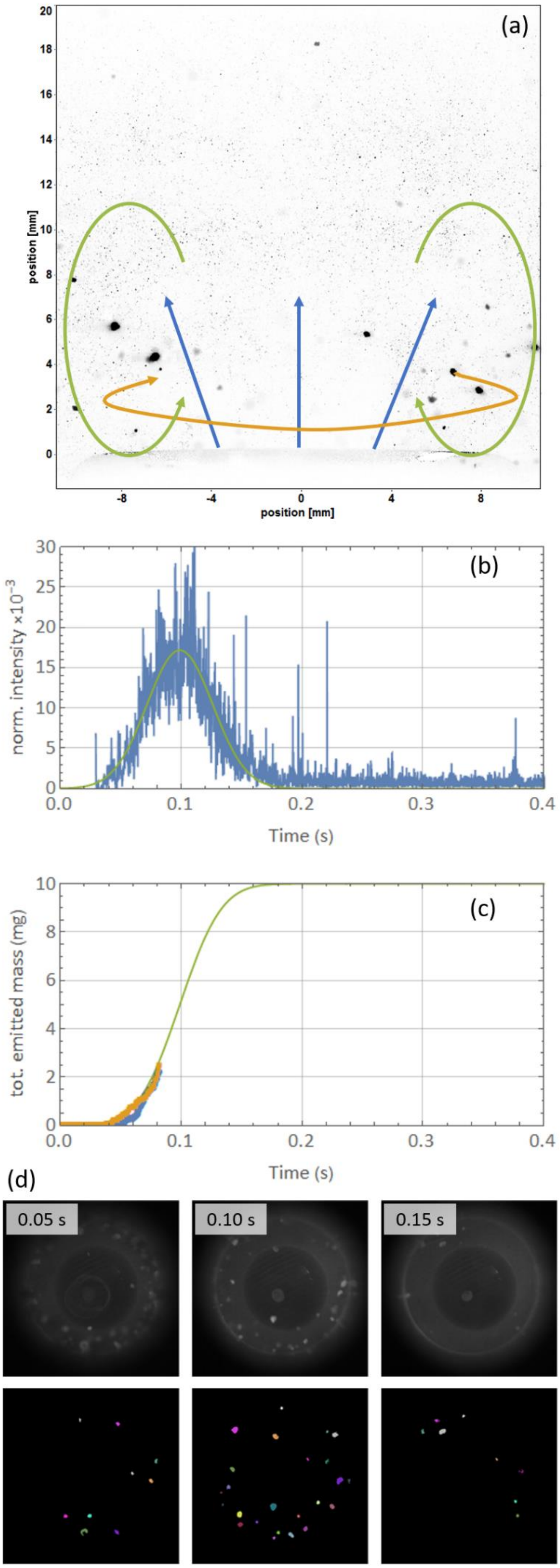

**Figure A.2: Image analysis of the top-view of the cup for the case $Q_{max} = 60$ l/min and $t_{ramp} = 0.3$ s.** (a) sketch of the recirculation motion of carrier particles already emitted from the device and trapped in the collection chamber. (b) Normalized photogram intensity as a function of time as obtained from the image analysis of the top-view movies of the inhaler cup. The green line represents a fit through a gaussian function. (c) carrier emitted mass as a function of time obtained as the cumulative distribution of the gaussian function on panel (b), the same data of Figure A.1 (d) are also reported for a comparison. (d) three photograms of the movie before, during and after the carrier emission peak, the bottom row images highlight the carrier particles detected by the image analysis algorithm.

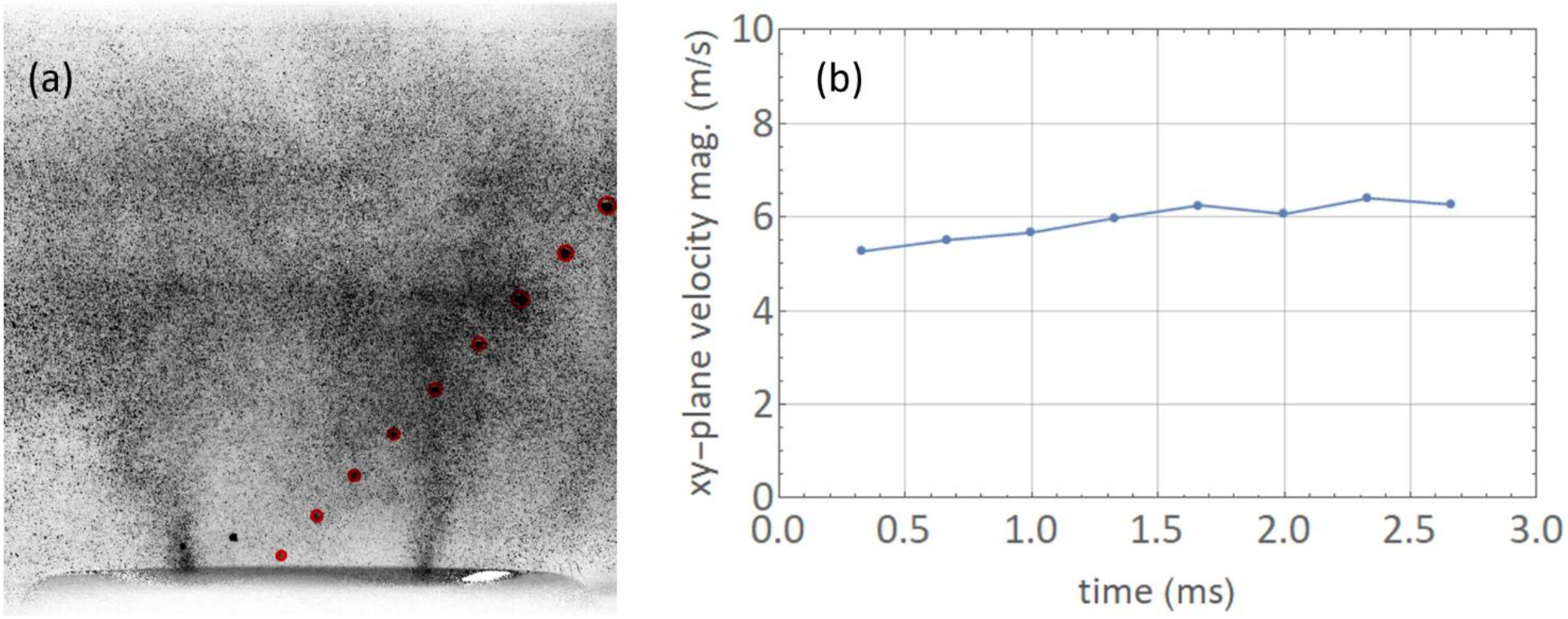

**Figure A.3: Carrier particle emission velocity**. (a) superimposition of 9 photograms showing the trajectory of a carrier particle, highlighted with the red circles. (b) calculated velocity for the particle of panel (a).

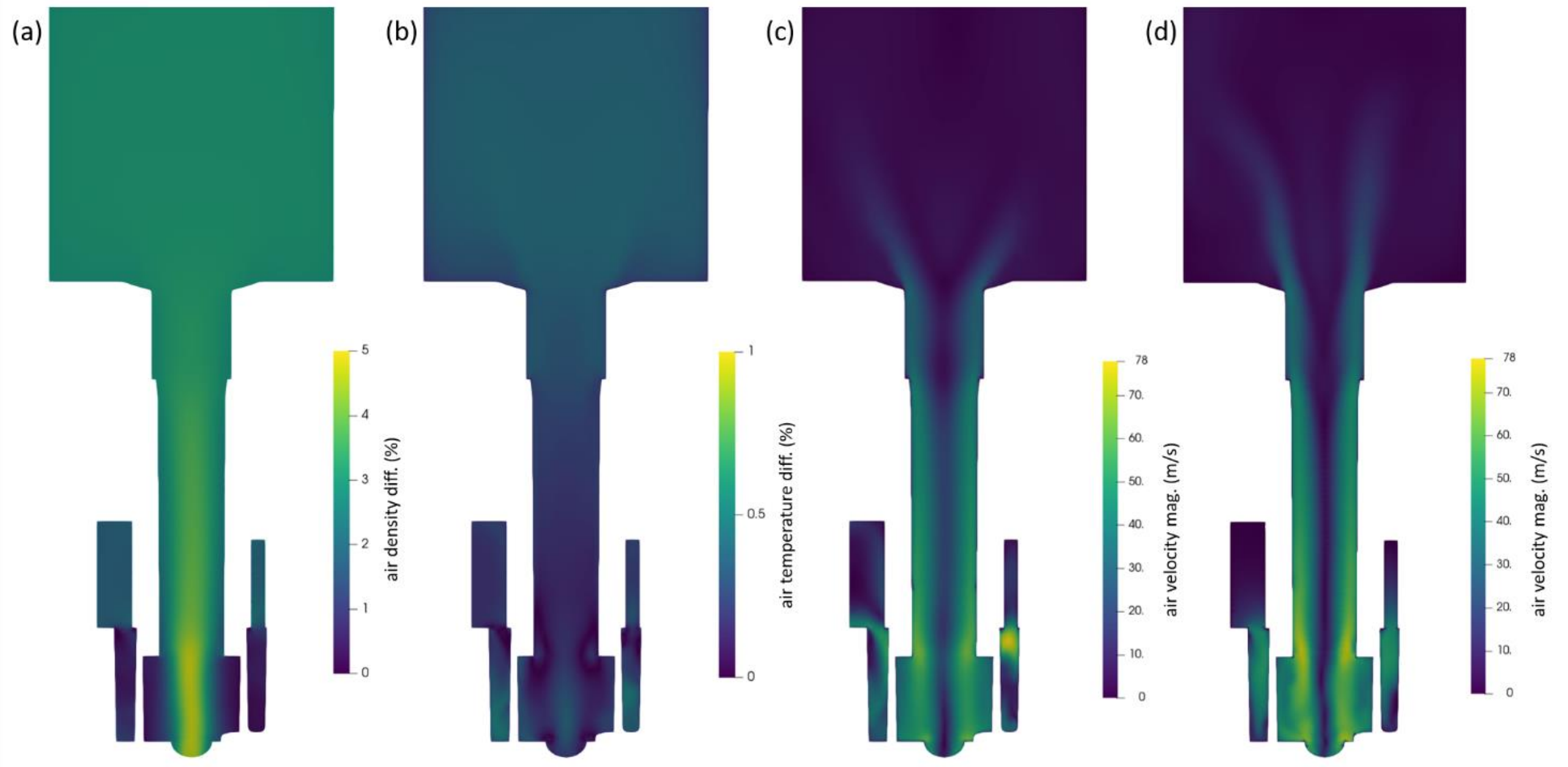

**Figure B.1: Comparison between compressible and incompressible flow assumption in the steady state condition for the case $Q_{max} = 60$ l/min**. (a) percent difference between the temperature field of the compressible flow solution and the reference constant temperature (23ºC) for the incompressible case. (b) percent difference between the density field of the compressible solution and the reference constant air density (see Table 3) for the incompressible case. (c) air velocity map from steady state compressible calculation. (d) instantaneous air velocity map from the incompressible time-dependent calculation once the steady state is reached.

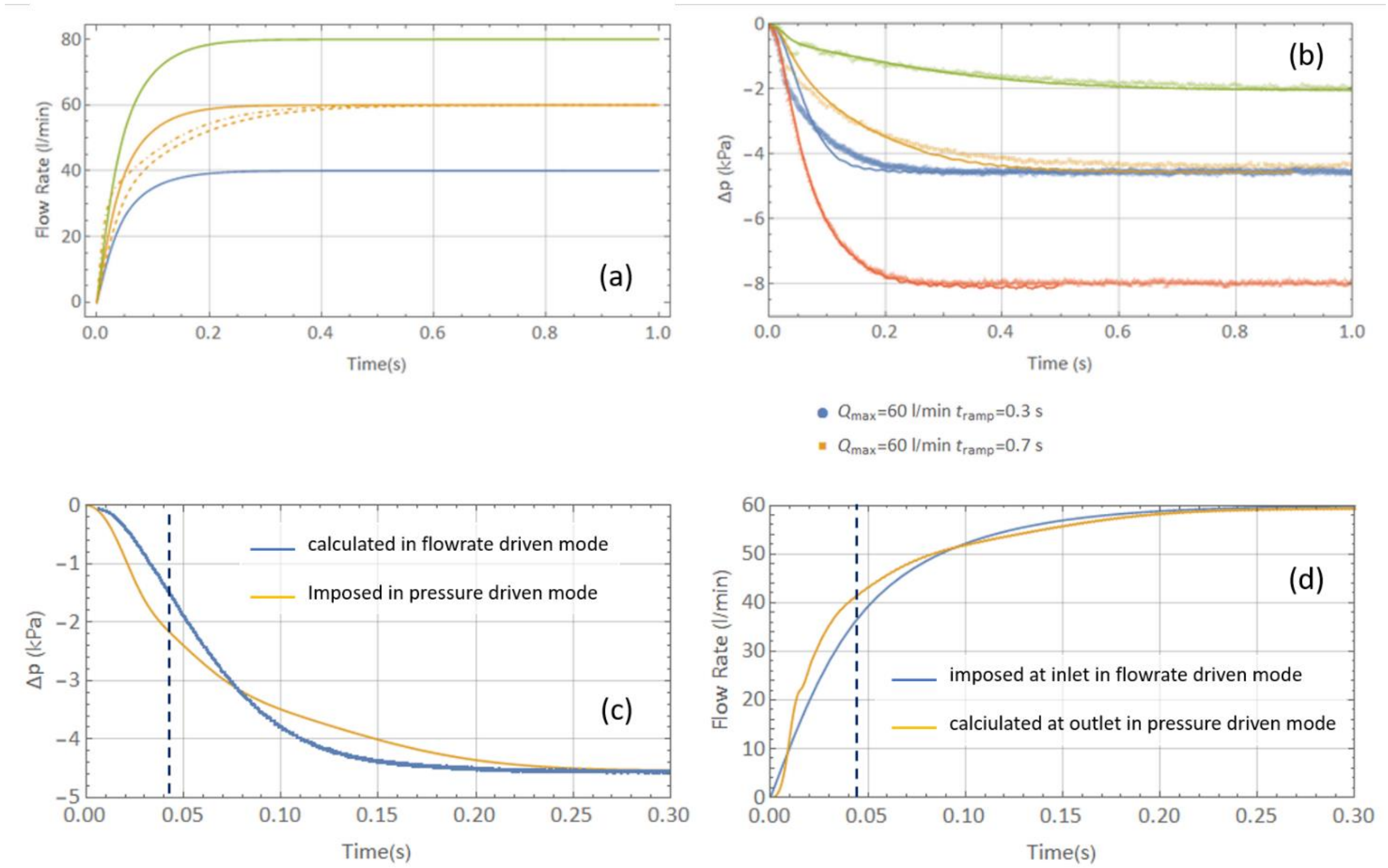

**Figure B.2: Pressure vs. flow driven modes**. (a) flow rate profiles fitted from PIV data and used as boundary conditions for the flow rate driven mode. (b) experimental (markers) and calculated (continuous lines) inlet/outlet pressure drops in flow rate driven mode. (c)-(d) Comparison between flow rate driven and pressure driven simulations, for the inlet/outlet pressure drop and flow rate, in the case $Q_{max} = 60$ l/min and $t_{ramp} = 0.3$ s. Black dashed lines represent the moment in which the dose protector opens upon reaching $\Delta p = 2$ kPa according the experimental observations of Pasquali et al. .

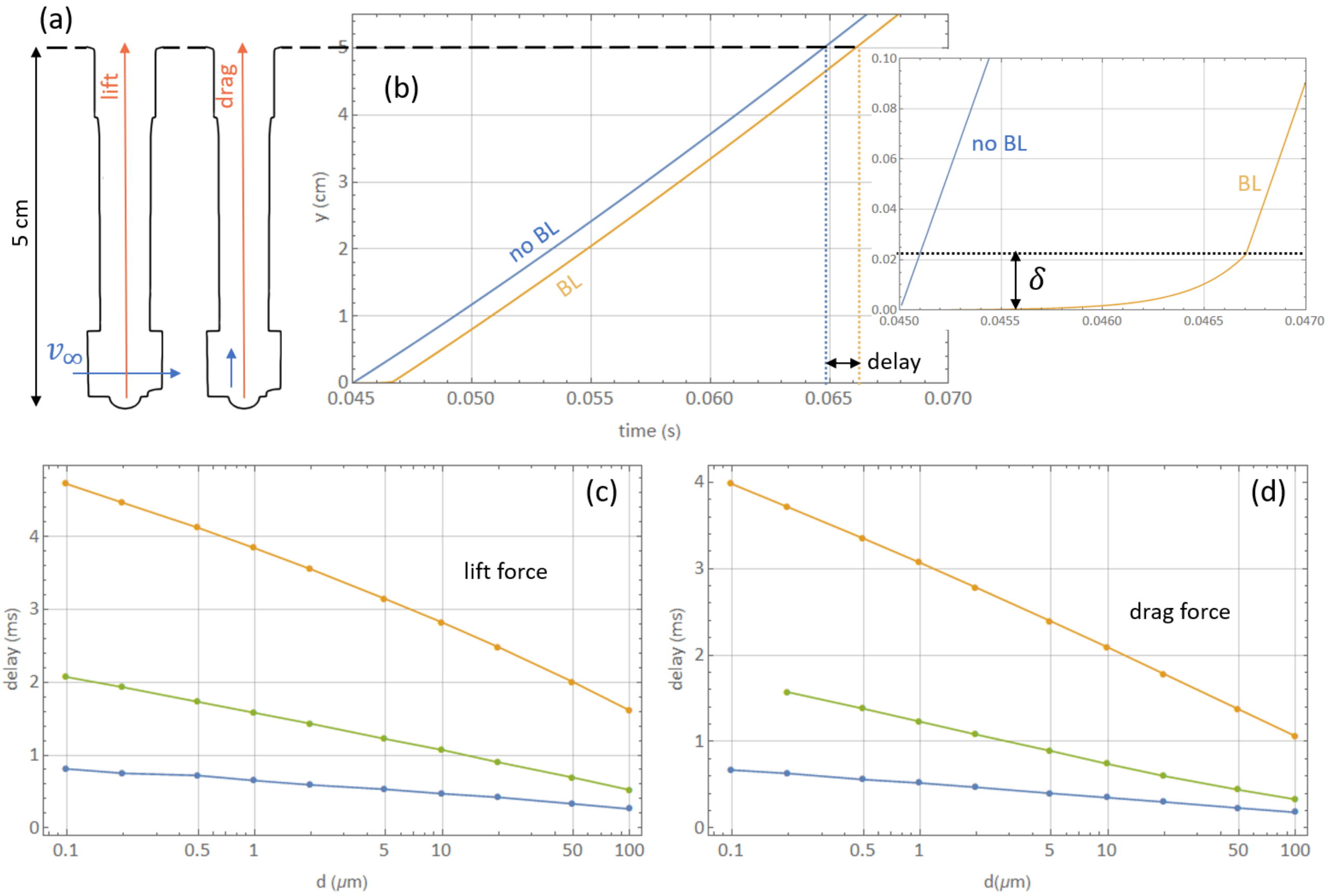

**Figure C.1: Importance of the boundary layer**. (a) sketch of the 1D model showing the direction of fluid velocity (blue arrows) and particle-fluid forces (red arrows). The air enters the swirl chamber from the two symmetrical inlets as shown schematically in Figure 9. (b) Particle position as a function of time according to the 1D model with and without the inclusion of the boundary layer, the inset magnifies the initial moments of the inhalation. A residence time can be estimated considering the total time necessary for the particles to reach the device outlet at 5 cm distance from the device bottom, including the boundary layer presence always results in a delay. (c) and (d) delay in the calculated residence time as a function of the API particle diameter for lift and drag forces respectively. No attenuation coefficient and no dose protector are applied for the blue curve, the attenuation coefficient is applied in the yellow curve, both the attenuation coefficient and the presence of the dose protector are included in for the green curve.

## Tables

| | pressure | velocity | $k$ | $\omega$ | $\nu_t$ |
|---|---|---|---|---|---|
| **Inlet** | fixedValue | inletOutlet | fixedValue | zeroGradient | zeroGradient |
| **Outlet** | $\Delta p(t)$ from eq.(A2) | pressureInletOutletVelocity | zeroGradient | zeroGradient | zeroGradient |
| **Walls** | zeroGradient | fixedValue | fixedValue | fixedValue | nutUSpaldingWallFunction |

Table 1: Boundary conditions for the CFD simulations in the openFoam jargon. While the meaning of zeroGradient and fixedValue is straightforward (for the specific fixed value see Table 2), the inletOutlet conditions are zero gradient conditions allowing for a flow reversal. nutUSpaldingWallFunction uses Spalding law to estimate the eddy viscosity at a given distance from the wall based on the fluid velocity in the first cell. Details can be found in the openFoam online manual [46].

| | pressure | velocity | $k$ | $\omega$ | $\nu_t$ |
|---|---|---|---|---|---|
| **Internal** | 0 | 0 | flat plate eqns. | flat plate eqns. | flat plate eqns. |
| **Inlet** | 0 | - | flat plate eqns. | - | - |
| **Outlet** | 0 | - | - | - | - |
| **Walls** | - | - | 0 | 0 | calculated |

Table 2: Initial conditions for the CFD simulations. The flat plate equations used to estimate the various initial fields are described in reference [52].

| - Constant | Meaning | p value | w value | p-w value | Units |
|---|---|---|---|---|---|
| $\rho_f$ | Air density | 1.16 | | | Kg/m$^3$ |
| $c$ | Speed of sound in air | 343 | | | m/s |
| $\mu$ | Air viscosity | $18.27 \times 10^{-6}$ | | | Pa s |
| $\rho_p$ | Lactose density | 1525 | | | Kg/m$^3$ |
| $Y$ | Young modulus | 5 | 10 | \ | MPa |
| $\nu$ | Poisson ratio | 0.45 | 0.45 | \ | dimensionless |
| $\mu_s$ | Sliding friction coefficient | $0.1 \div 0.8$ | $0.1 \div 0.8$ | $0.1 \div 0.8$ | dimensionless |
| $\mu_r$ | Rolling friction coefficient | $0.1 \div 0.8$ | $0.1 \div 0.8$ | $0.1 \div 0.8$ | dimensionless |
| $e$ | Restitution coefficient | $0.1 \div 0.8$ | $0.1 \div 0.8$ | $0.1 \div 0.8$ | dimensionless |

Table 3: Material parameters entering both the CFD and DEM equations. In the DEM case some of the parameters have been split into particle (p) or wall (w) values, representing intrinsic powder or wall properties or parameters ruling the particle-particle interaction. The particle-wall values (p-w) rule the particle-walls interactions.

| Parameter | Range | Explanation |
|---|---|---|
| $I_{tr}$ | 0.3 and 0.7 | Determines if a pixel is perceived as part of a particle or part of the image background |
| Equivalent radius | > 100 µm<br>< 1600 µm | Filters too small API aggregates or too large cluster resulting from the superimposition or two particles. |
| Elongation | < 0.5 | Filters spurious clusters extremely elongated in one direction. |
| Convex Hull perimeter/ cluster perimeter | > 0.6 | Filters spurious clusters whose perimeter is too irregular compared to real carrier particles. |

Table A.1: filtering conditions used to eliminate spurious clusters and artifacts of the image binarization.

| $Q_{max}$(l/min) | $t_{ramp}(s)$ | $p_0$(kPa) | $a_1$ | $a_2$ | $a_3$ |
|---|---|---|---|---|---|
| **60** | 0.3 | 1.51 | 6.22 | 39.82 | 247.66 |
| **60** | 0.7 | 1.46 | 8.55 | 69.89 | 1448.78 |
| **80** | 0.3 | 2.64 | 6.98 | 24.10 | 107.20 |
| **40** | 1.2 | 0.67 | 7.06 | 67.34 | 2932.65 |

Table B.1: fitting parameters for equation (A2) applied to the 4 pressure drops profiles of Figure 3 (b).

| Symbol | Meaning | Units |
|---|---|---|
| $\rho_p$ | Particle density | Kg/m$^3$ |
| $\mu_s$ | Sliding friction coefficient | dimensionless |
| $\mu_r$ | Rolling friction coefficient | dimensionless |
| $e$ | Restitution coefficient | dimensionless |
| $\rho_f$ | Fluid density | Kg/m$^3$ |
| $\mu$ | Fluid dynamic viscosity | Pa s |
| $\nu_t$ | Kinematic eddy viscosity | m$^2$/s |
| $\omega$ | Specific rate of turbulent energy dissipation | 1/s |
| $k$ | Turbulence kinetic energy per unit mass | m$^2$/s$^2$ |
| $c$ | Speed of sound in air | m/s |
| $Ma$ | Mach number | dimensionless |
| $t_{ramp}$ | Ramping time for the applied flow rate | s |
| $t_{tot}$ | Total duration of the flow rate application | s |
| $Q(t)$ | Air flow rate applied to the device | l/min or Nm$^3$/s |
| $Q_{max}$ | Maximum air flow rate applied to the device | l/min or Nm$^3$/s |
| $\overrightarrow{v^f}$ | Fluid velocity vector field | m/s |
| $\overrightarrow{v^p}$ | Individual particle velocity | m/s |
| $\overrightarrow{v_{slip}}$ | Particle-fluid slip velocity | m/s |
| $v_{max}^f$ | Maximum fluid velocity obtained from simulations | m/s |
| $v_{max}^p$ | Maximum particle velocity obtained from simulations | m/s |
| $v_{max}^{slip}$ | Maximum slip velocity obtained from simulations | m/s |
| $v_\infty$ | Fluid free stream velocity (velocity far from surfaces) | m/s |
| $D$ | Pipe diameter | m or mm |
| $\ell$ | Inlet width | m or mm |
| $Re^f$ | Fluid Reynolds number | dimensionless |
| $Re^p$ | Particle Reynolds number | dimensionless |
| $Co^f$ | Fluid Courant number | dimensionless |
| $Co^p$ | Particle Courant number | dimensionless |
| $Stk$ | Stokes number | dimensionless |
| $n$ | Fluid volume fraction | dimensionless |
| $d$ | Particle diameter | µm or m |
| $\Delta t_{DEM}$ | Time integration step for DEM calculations | s |
| $\Delta t_{CFD}$ | Time integration step for CFD calculations | s |
| $\Delta t_{LT}$ | Time integration step for lagrangian tracing | s |
| $\Delta x$ | Spatial discretization step for CFD calculations | µm or m |
| $\overline{\Delta x}$ | Average spatial discretization step for CFD calculations | µm or m |
| $\Delta x_{min}$ | Minimum spatial discretization step for CFD calculations | µm or m |
| $\tau_f$ | Fluid relaxation time | s |
| $\tau_p$ | Particle relaxation time | s |
| $\tau^*$ | Inhaler relaxation time | s |
| $Re_x$ | Flat plate local Reynolds number | dimensionless |
| $\delta$ | Boundary layer thickness | µm or m |
| $y^+$ | Dimensionless distance from the walls | dimensionless |

Table 4: List of symbols.

## Bibliography

[1] W.H. Finlay, The mechanics of inhaled pharmaceutical aerosols, Academic Press, 2001.

[2] M. Hoppentocht, P. Hagedoorn, H.W. Frijlink, A.H. de Boer, Technological and practical challenges of dry powder inhalers and formulations, Adv. Drug Deliv. Rev. 75 (2014) 18–31. doi:10.1016/j.addr.2014.04.004.

[3] D. Traini, P.M. Young, Formulation of Inhalation Medicines, in: Inhal. Drug Deliv., John Wiley & Sons, Ltd, 2013: pp. 31–45. doi:10.1002/9781118397145.ch3.

[4] K. Berkenfeld, A. Lamprecht, J.T. McConville, Devices for Dry Powder Drug Delivery to the Lung, AAPS PharmSciTech. 16 (2015) 479–490. doi:10.1208/s12249-015-0317-x.

[5] A.H. de Boer, P. Hagedoorn, M. Hoppentocht, F. Buttini, F. Grasmeijer, H.W. Frijlink, Dry powder inhalation: past, present and future, Expert Opin. Drug Deliv. 14 (2017) 499–512. doi:10.1080/17425247.2016.1224846.

[6] A.H. De Boer, H.K. Chan, R. Price, A critical view on lactose-based drug formulation and device studies for dry powder inhalation: Which are relevant and what interactions to expect?, Adv. Drug Deliv. Rev. 64 (2012) 257–274. doi:10.1016/j.addr.2011.04.004.

[7] A.M. Healy, M.I. Amaro, K.J. Paluch, L. Tajber, Dry powders for oral inhalation free of lactose carrier particles, Adv. Drug Deliv. Rev. 75 (2014) 32–52. doi:10.1016/j.addr.2014.04.005.

[8] F. Buttini, G. Colombo, P.C.L. Kwok, W.T. Wui, Aerodynamic Assessment for Inhalation Products: Fundamentals and Current Pharmacopoeial Methods, in: Inhal. Drug Deliv., John Wiley & Sons, Ltd, 2013: pp. 91–119. doi:10.1002/9781118397145.ch6.

[9] P.M. Young, O. Wood, J. Ooi, D. Traini, The influence of drug loading on formulation structure and aerosol performance in carrier based dry powder inhalers, Int. J. Pharm. 416 (2011) 129–135. doi:10.1016/j.ijpharm.2011.06.020.

[10] G. Pilcer, N. Wauthoz, K. Amighi, Lactose characteristics and the generation of the aerosol, Adv.

Drug Deliv. Rev. 64 (2012) 233–256. doi:10.1016/j.addr.2011.05.003.

[11] J. Shur, R. Price, D. Lewis, P.M. Young, G. Woollam, D. Singh, S. Edge, From single excipients to dual excipient platforms in dry powder inhaler products, Int. J. Pharm. 514 (2016) 374–383. doi:https://doi.org/10.1016/j.ijpharm.2016.05.057.

[12] P. Du, J. Du, H.D.C. Smyth, Evaluation of Granulated Lactose as a Carrier for DPI Formulations 1: Effect of Granule Size, Ageing Int. 15 (2014) 1417–1428. doi:10.1208/s12249-014-0166-z.

[13] J. Rudén, A. Vajdi, G. Frenning, T. Bramer, Comparison Between High and Low Shear Mixers in the Formation of Adhesive Mixtures for Dry Powder Inhalers, Respir. Drug Deliv. Eur. 2019. 2 (2019) 457–460.

[14] W. Kaialy, On the effects of blending, physicochemical properties, and their interactions on the performance of carrier-based dry powders for inhalation-A review, Adv. Colloid Interface Sci. 235 (2016) 70–89. doi:10.1016/j.cis.2016.05.014.

[15] M. Hertel, E. Schwarz, M. Kobler, S. Hauptstein, H. Steckel, R. Scherließ, The influence of high shear mixing on ternary dry powder inhaler formulations, Int. J. Pharm. 534 (2017) 242–250. doi:10.1016/j.ijpharm.2017.10.033.

[16] M. Djokić, K. Kachrimanis, L. Solomun, J. Djuriš, D. Vasiljević, S. Ibrić, A study of jet-milling and spray-drying process for the physicochemical and aerodynamic dispersion properties of amiloride HCl, Powder Technol. 262 (2014) 170–176. doi:10.1016/j.powtec.2014.04.066.

[17] S. Yeung, D. Traini, A. Tweedie, D. Lewis, T. Church, P.M. Young, Effect of Dosing Cup Size on the Aerosol Performance of High-Dose Carrier-Based Formulations in a Novel Dry Powder Inhaler, J. Pharm. Sci. 108 (2019) 949–959. doi:10.1016/j.xphs.2018.09.033.

[18] T. Kopsch, D. Murnane, D. Symons, Optimizing the Entrainment Geometry of a Dry Powder Inhaler: Methodology and Preliminary Results, Pharm. Res. 33 (2016) 2668–2679. doi:10.1007/s11095-016-1992-3.

[19] M. Nakhaei, B. Lu, Y. Tian, W. Wang, K. Dam-Johansen, H. Wu, CFD modeling of gas-solid cyclone separators at ambient and elevated temperatures, Processes. 8 (2020) 1–26. doi:10.3390/pr8020228.

[20] S. Bnà, R. Ponzini, M. Cestari, C. Cavazzoni, C. Cottini, A. Benassi, Investigation of particle dynamics and classification mechanism in a spiral jet mill through computational fluid dynamics and discrete element methods, Powder Technol. 364 (2020) 746–773. doi:10.1016/j.powtec.2020.02.029.

[21] M. Sommerfeld, Y. Cui, S. Schmalfuß, Potential and constraints for the application of CFD combined with Lagrangian particle tracking to dry powder inhalers, Eur. J. Pharm. Sci. 128 (2019) 299–324. doi:10.1016/j.ejps.2018.12.008.

[22] J. Tu, K. Inthavong, G. Ahmadi, Computational Fluid and Particle Dynamics in the Human Respiratory System, Springer Netherlands, 2013.

[23] W. Wong, D.F. Fletcher, D. Traini, H.K. Chan, P.M. Young, The use of computational approaches in inhaler development, Adv. Drug Deliv. Rev. 64 (2012) 312–322. doi:10.1016/j.addr.2011.10.004.

[24] C.A. Ruzycki, E. Javaheri, W.H. Finlay, The use of computational fluid dynamics in inhaler design, Expert Opin. Drug Deliv. 10 (2013) 307–323. doi:10.1517/17425247.2013.753053.

[25] T. Suwandecha, W. Wongpoowarak, T. Srichana, Computer-aided design of dry powder inhalers using computational fluid dynamics to assess performance, Pharm. Dev. Technol. 21 (2016) 54–60. doi:10.3109/10837450.2014.965325.

[26] T. Kopsch, D. Murnane, D. Symons, A personalized medicine approach to the design of dry powder inhalers: Selecting the optimal amount of bypass, Int. J. Pharm. 529 (2017) 589–596. doi:10.1016/j.ijpharm.2017.07.002.

[27] M. Ariane, M. Sommerfeld, A. Alexiadis, Wall collision and drug-carrier detachment in dry powder inhalers: Using DEM to devise a sub-scale model for CFD calculations, Powder Technol. 334 (2018) 65–75. doi:10.1016/j.powtec.2018.04.051.

[28] J. Yang, C.Y. Wu, M. Adams, DEM analysis of the effect of particle-wall impact on the dispersion performance in carrier-based dry powder inhalers, Int. J. Pharm. 487 (2015) 32–38. doi:10.1016/j.ijpharm.2015.04.006.

[29] Z.B. Tong, R.Y. Yang, K.W. Chu, A.B. Yu, S. Adi, H.K. Chan, Numerical study of the effects of particle size and polydispersity on the agglomerate dispersion in a cyclonic flow, Chem. Eng. J. 164 (2010) 432–441. doi:https://doi.org/10.1016/j.cej.2009.11.027.

[30] Z.B. Tong, S. Adi, R.Y. Yang, H.K. Chan, A.B. Yu, Numerical investigation of the de-agglomeration mechanisms of fine powders on mechanical impaction, J. Aerosol Sci. 42 (2011) 811–819. doi:https://doi.org/10.1016/j.jaerosci.2011.07.004.

[31] Z. Tong, W. Zhong, A. Yu, H.K. Chan, R. Yang, CFD-DEM investigation of the effect of agglomerate-agglomerate collision on dry powder aerosolisation, J. Aerosol Sci. 92 (2016) 109–121. doi:10.1016/j.jaerosci.2015.11.005.

[32] Y. Cui, M. Sommerfeld, Forces on micron-sized particles randomly distributed on the surface of larger particles and possibility of detachment, Int. J. Multiph. Flow. 72 (2015) 39–52. doi:https://doi.org/10.1016/j.ijmultiphaseflow.2015.01.006.

[33] D. Nguyen, J. Remmelgas, I.N. Björn, B. van Wachem, K. Thalberg, Towards quantitative prediction of the performance of dry powder inhalers by multi-scale simulations and experiments, Int. J. Pharm. 547 (2018) 31–43. doi:10.1016/j.ijpharm.2018.05.047.

[34] Z. Tong, H. Kamiya, A. Yu, H.K. Chan, R. Yang, Multi-scale modelling of powder dispersion in a carrier-based inhalation system, Pharm. Res. 32 (2015) 2086–2096. doi:10.1007/s11095-014-1601-2.

[35] Z. Qiao, Z. Wang, C. Zhang, S. Yuan, Y. Zhu, J. Wang, Simulation of Dry Powder Inhalers: Combining Micro-Scale, Meso-Scale and Macro-Scale Modeling, AIChE J. 59 (2012) 215–228. doi:10.1002/aic.

[36] M. Casey, T. Wintergerste, Special interest group on quality and trust in industrial CFD, best practice guidelines, version 1. ERCOFTAC Qinetiq; Farnborough, Hampshire, UK: 2000, Farnborough,

Hampshire, UK, 2000.

[37] M. Sommerfeld, B. van Wachem, R. Oliemans, Best practice guidelines for computational fluid dynamics of dispersed multiphase flows, ERCOFTAC SIAMUF, Gothenburg, Sweden, 2008.

[38] Assessing Credibility of Computational Modeling through Verification and Validation: Application to Medical Devices, V&V 40, ASME, New York, NY, 2018.

[39] I. Pasquali, C. Merusi, G. Brambilla, E.J. Long, G.K. Hargrave, H.K. Versteeg, Optical diagnostics study of air flow and powder fluidisation in Nexthaler® - Part I: Studies with lactose placebo formulation, Int. J. Pharm. 496 (2015) 780–791. doi:10.1016/j.ijpharm.2015.10.072.

[40] C. Merusi, G. Brambilla, E.J. Long, G.K. Hargrave, H.K. Versteeg, Optical diagnostics studies of air flow and powder fluidisation in Nexthaler®. Part II: Use of fluorescent imaging to characterise transient release of fines from a dry powder inhaler, Int. J. Pharm. 549 (2018) 96–108. doi:10.1016/j.ijpharm.2018.07.032.

[41] M. Corradi, H. Chrystyn, B.G. Cosio, M. Pirozynski, S. Loukides, R. Louis, M. Spinola, O.S. Usmani, NEXThaler, an innovative dry powder inhaler delivering an extrafine fixed combination of beclometasone and formoterol to treat large and small airways in asthma, Expert Opin. Drug Deliv. 11 (2014) 1497–1506. doi:10.1517/17425247.2014.928282.

[42] S. Yeung, D. Traini, A. Tweedie, D. Lewis, T. Church, P.M. Young, Assessing Aerosol Performance of a Dry Powder Carrier Formulation with Increasing Doses Using a Novel Inhaler, AAPS PharmSciTech. 20 (2019) 1–12. doi:10.1208/s12249-019-1302-6.

[43] A. Hager, C. Kloss, S. Pirker, C. Goniva, Parallel Resolved Open Source CFD-DEM: Method, Validation and Application, J. Comput. Multiph. Flows. 6 (2014) 13–27. doi:10.1260/1757-482X.6.1.13.

[44] C. Goniva, C. Kloss, N. Deen, H. Kuipers, S. Pirker, Influence of rolling friction on single spout fluidized bed simulation, Particuology. 10 (2012) 582–591. doi:10.1016/j.partic.2012.05.002.

[45] DCS Computing, (n.d.). https://www.aspherix-dem.com/.

[46] The OpenFOAM Foundation, (n.d.). https://openfoam.org/.

[47] H.G. Weller, G. Tabor, H. Jasak, C. Fureby, A tensorial approach to computational continuum mechanics using object-oriented techniques, Comput. Phys. 12 (1998) 620–631.

[48] C. Kloss, C. Goniva, A. Hager, S. Amberger, S. Pirker, Models, algorithms and validation for opensource DEM and CFD-DEM, Prog. Comput. Fluid Dyn. An Int. J. 12 (2012) 140. doi:10.1504/PCFD.2012.047457.

[49] Y. Cengel, J. Cimbala, Fluid Mechanics: Fundamentals and Applications, McGraw Hill Education, 2018.

[50] F.R. Menter, Two-equation eddy-viscosity turbulence models for engineering applications, AIAA J. 32 (1994) 1598–1605. doi:10.2514/3.12149.

[51] S.M. Salim, S.C. Cheah, Wall y+ Strategy for Dealing with Wall-bounded Turbulent Flows Vol II, Proc. Int. MultiConference Eng. Comput. Sci. 2009. (2009).

[52] H. Schlichting, K. Gersten, Boundary Layer Theory, Springer-Verlag, 2017.

[53] H.R. Norouzi, R. Zarghami, R. Sotudeh-Gharebagh, N. Mostoufi, Coupled CFD-DEM Modeling, John Wiley & Sons, 2016.

[54] M. Paulick, M. Morgeneyer, A. Kwade, Review on the influence of elastic particle properties on DEM simulation results, Powder Technol. 283 (2015) 66–76. doi:10.1016/j.powtec.2015.03.040.

[55] T. Brosh, H. Kalman, A. Levy, Accelerating CFD-DEM simulation of processes with wide particle size distributions, Particuology. 12 (2014) 113–121. doi:10.1016/j.partic.2013.04.008.

[56] A. Dehbi, Turbulent particle dispersion in arbitrary wall-bounded geometries: A coupled CFD-Langevin-equation based approach, Int. J. Multiph. Flow. 34 (2008) 819–828. doi:https://doi.org/10.1016/j.ijmultiphaseflow.2008.03.001.

[57] A. Di Renzo, F.P. Di Maio, Comparison of contact-force models for the simulation of collisions in

DEM-based granular flow codes, Chem. Eng. Sci. 59 (2004) 525–541. doi:https://doi.org/10.1016/j.ces.2003.09.037.

[58] J. Ai, J.-F. Chen, J.M. Rotter, J.Y. Ooi, Assessment of rolling resistance models in discrete element simulations, Powder Technol. 206 (2011) 269–282. doi:https://doi.org/10.1016/j.powtec.2010.09.030.

[59] E. Barthel, Adhesive elastic contacts: JKR and more, J. Phys. D. Appl. Phys. 41 (2008) 163001. doi:10.1088/0022-3727/41/16/163001.

[60] Z.Y. Zhou, S.B. Kuang, K.W. Chu, A.B. Yu, Discrete particle simulation of particle-fluid flow: Model formulations and their applicability, J. Fluid Mech. 661 (2010) 482–510. doi:10.1017/S002211201000306X.

[61] H.P. Zhu, Z.Y. Zhou, R.Y. Yang, A.B. Yu, Discrete particle simulation of particulate systems: Theoretical developments, Chem. Eng. Sci. 62 (2007) 3378–3396. doi:10.1016/j.ces.2006.12.089.

[62] R.J. Koch, D.L. Hill, INERTIAL EFFECTS IN SUSPENSION AND POROUS-MEDIA FLOWS, Annu. Rev. Fluid Mech. 33 (2001) 619–647.

[63] A. Haider, O. Levenspiel, Drag coefficient and terminal velocity of spherical and nonspherical particles, Powder Technol. 58 (1989) 63–70. doi:https://doi.org/10.1016/0032-5910(89)80008-7.

[64] L.G. Sweeney, W.H. Finlay, Lift and drag forces on a sphere attached to a wall in a Blasius boundary layer, J. Aerosol Sci. 38 (2007) 131–135. doi:10.1016/j.jaerosci.2006.09.006.

[65] J.B. Mclaughlin, Inertial migration of a small sphere in linear shear flows, J. Fluid Mech. 224 (1991) 261–274. doi:10.1017/S0022112091001751.

[66] E. Loth, A.J. Dorgan, An equation of motion for particles of finite Reynolds number and size, Environ. Fluid Mech. 9 (2009) 187–206. doi:10.1007/s10652-009-9123-x.

[67] P. Cherukat, J.B. McLaughlin, A.L. Graham, The inertial lift on a rigid sphere translating in a linear

shear flow field, Int. J. Multiph. Flow. 20 (1994) 339–353. doi:https://doi.org/10.1016/0301-9322(94)90086-8.

[68] D. Leighton, A. Acrivos, The lift on a small sphere touching a plane in the presence of a simple shear flow, Z. Angew. Math. Phys. 36 (1985) 174–178. doi:https://doi.org/10.1007/BF00949042.

[69] C.M. Boyce, D.J. Holland, S.A. Scott, J.S. Dennis, Limitations on Fluid Grid Sizing for Using Volume-Averaged Fluid Equations in Discrete Element Models of Fluidized Beds, Ind. Eng. Chem. Res. 54 (2015) 10684–10697. doi:10.1021/acs.iecr.5b03186.

[70] S. Pirker, D. Kahrimanovic, C. Goniva, Improving the applicability of discrete phase simulations by smoothening their exchange fields, Appl. Math. Model. 35 (2011) 2479–2488. doi:10.1016/j.apm.2010.11.066.

[71] S. Radl, B. Capa Gonzales, C. Goniva, S. Pirker, State of the Art in Mapping Schemes for Dilute and Dense Euler-Lagrange Simulations, in: Prog. Appl. CFD, SINTEF Academic Press, 2015: pp. 103–112.

[72] A.D. Gosman, E. loannides, Aspects of Computer Simulation of Liquid-Fueled Combustors, J. Energy. 7 (1983) 482–490. doi:10.2514/3.62687.

[73] A. Tweedie, F. Mason, D. Lewis, Investigating the Effect of Modified Breath Actuated Mechanisms on the Dispersion Performance of the NEXThaler®, Drug Deliv. to Lungs 26. (2015).

[74] A. Tweedie, D. Lewis, Enhancing the Performance of Dry Powder Inhalers: Breath Actuated Mechanisms, ONdrugDelivery Mag. (2016) 34–38.

[75] B.H. Kaye, Particle Shape Characterization, in: M.E. Fayed, L. Otten (Eds.), Handb. Powder Sci. Technol., Springer, Boston, MA, 1997. doi:https://doi.org/10.1007/978-1-4615-6373-0_2.

[76] A.H. De Boer, D. Gjaltema, P. Hagedoorn, M. Schaller, W. Witt, H.W. Frijlink, Design and application of a new modular adapter for laser diffraction characterization of inhalation aerosols, Int. J. Pharm. 249 (2002) 233–245. doi:10.1016/S0378-5173(02)00527-6.

[77] K. Bass, D. Farkas, W. Longest, Optimizing Aerosolization Using Computational Fluid Dynamics in a Pediatric Air-Jet Dry Powder Inhaler, AAPS PharmSciTech. 20 (2019) 1–19. doi:10.1208/s12249-019-1535-4.

[78] J. Tibbatts, P. Mendes, P. Villax, Understanding the Power Requirements for Efficient Dispersion in Powder Inhalers: Comparing CFD Predictions and Experimental Measurements, Respir. Drug Deliv. 2010. 1 (2010) 323–330.

[79] S. Yeung, D. Traini, D. Lewis, P.M. Young, Dosing challenges in respiratory therapies, Int. J. Pharm. 548 (2018) 659–671. doi:https://doi.org/10.1016/j.ijpharm.2018.07.007.

[80] P.M. Young, A. Tweedie, D. Lewis, T. Church, D. Traini, Limitations of high dose carrier based formulations, Int. J. Pharm. 544 (2018) 141–152. doi:10.1016/j.ijpharm.2018.04.012.

[81] P.W. Longest, L.T. Holbrook, In silico models of aerosol delivery to the respiratory tract - Development and applications, Adv. Drug Deliv. Rev. 64 (2012) 296–311. doi:10.1016/j.addr.2011.05.009.

[82] Y. Feng, Z. Xu, A. Haghnegahdar, Computational Fluid-Particle Dynamics Modeling for Unconventional Inhaled Aerosols in Human Respiratory Systems, in: K. Volkov (Ed.), Aerosols - Sci. Case Stud., IntechOpen, London, UK, 2016. doi:10.5772/65361.

[83] C.-L. Lin, M.H. Tawhai, E.A. Hoffman, Multiscale image-based modeling and simulation of gas flow and particle transport in the human lungs, Interdiscip Rev Syst Biol Med. 5 (2013) 643. doi:10.1038/jid.2014.371.

[84] M.B. Dolovich, A. Kuttler, T.J. Dimke, O.S. Usmani, Biophysical model to predict lung delivery from a dual bronchodilator dry-powder inhaler, Int. J. Pharm. X. 1 (2019) 100018. doi:10.1016/j.ijpx.2019.100018.

[85] A.H. De Boer, D. Gjaltema, P. Hagedoorn, H.W. Frijlink, Can “extrafine” dry powder aerosols improve lung deposition?, Eur. J. Pharm. Biopharm. 96 (2015) 143–151. doi:10.1016/j.ejpb.2015.07.016.